# Fully-chromophoric ferroelectric nematics for electronic electro-optics

*Xi Chen[1], Andrew Bradfield[1], Chirag Patel[2], Pavel Savechenkov[2], Jason W. Sickler[2], Gianlorenzo Masini[2], Joseph E. Maclennan[1], Noel A. Clark[1], Cory Pecinovsky[2]\*, Matthew A. Glaser[1]\**

[1]Department of Physics, University of Colorado Boulder, Boulder, CO 80309

[2]Polaris Electro-Optics, Inc., Broomfield, CO 80020

\*corresponding authors: matthew.glaser@colorado.edu, cory@polariseo.com

## Abstract

Electronic electro-optic (EEO) phase modulation is a key emerging technology for the chip-scale interconversion of signals between the electronic and photonic domains. The recent discovery of the ferroelectric nematic ($N_F$) liquid crystal phase, a three dimensional fluid of rod-shaped organic molecules having near-perfect equilibrium polar molecular orientational order, offers attractive opportunities for the creation of second-order nonlinear optical materials for EEO. Here we propose and realize a design motif for $N_F$ EEO molecules in which few-nanometer-long molecular rods are functionalized both for electrostatic end-to-end association, facilitating $N_F$ phase formation, and for chromophoric optical nonlinearity, enabling high EEO efficiency, a combination enabling an active second-order nonlinear EEO medium that is 100% chromophoric.

Introduction

Electronic electro-optic (EEO) phase modulation, where the induced phase change of an optical electric field is proportional to an applied DC or AC electric field (the Pockels effect [1]) is a key emerging second order nonlinear optical technology for the chip-scale interconversion of signals between the electronic and photonic domains.  The high speeds desired, with anticipated device frequencies ranging into the terahertz regime [2], require the exploitation of field-driven electron motion in macroscopically non-centrosymmetric media [3].  The classic Pockels development, and currently the most broadly applied linear EEO devices, are based on ferroelectric crystals, in which unit cells with dipolar relative displacement of positive and negatively charged atoms produce macroscopic polarization.  Modulation of such a charge separation with applied external electric field *E* generates a proportional change in optical refractive index *Δn*, and thereby optical phase modulation.  The crystalline ferroelectric LiNbO$_3$ is currently well-established as the EEO material of choice, combining attractive chemical and optical properties with a substantial EEO coefficient $Δn/E \propto r_{eff} = 32\text{pm/V}$ [1].

More broadly, a variety of soft matter polar materials systems are being explored and developed, which employ the optical nonlinearity of conjugated *π*-electron systems in organic molecules for advanced EEO.  Molecular *π*-electron systems are fast and are highly nonlinear, some with EEO coefficients many orders of magnitude larger than those of LiNbO$_3$ [2] [3].  Organic molecules also offer an essentially unlimited potential for structural variation, a key advantage used, for example, in the development of the materials platforms of LCD and OLED display technologies, in which hundreds of thousands of new molecules were made [4].  Thus, molecular organic EEO has seen a steady increase of performance figures of merit over the past few decades [5, 6], and nonlinear organic materials have been integrated into a variety of platforms and device structures, enabling high-speed modulators with attractive performance parameters and data rates approaching the terahertz regime [3].

Obtaining uniform, stable polar alignment order is a key goal in the development of such organic Pockels applications. A active media for molecular organic EEO phase modulation have been typically made by incorporating the active conjugated dipolar molecules as dopants into organic phases which are apolar in equilibrium.  The required macroscopic polar ordering is then achieved in a poling process by a combination of temperature cycling between fluid and glassy states with an applied DC electric field application, using polymer thermosets as hosts, for example [7, 8].  This strategy has enjoyed some success but imposes significant constraints on component design, typically requiring device geometries that can apply high electric field, and requiring EEO-active components that are at the same time chemically stable, dipolar, and functionalized: (i) for large *r*-values, *i.e.* for large first molecular hyperpolarizability *β*; (ii) for solubility in the host material; and (iii) for spontaneous distribution in the host material in a fashion optimized for cooperative polar ordering [7, 8].

We advance molecular organic EEO phase modulation by incorporating active conjugated dipolar organic molecules in, the ferroelectric nematic (N$_F$) liquid crystal phase [9], which brings transformative features to EEO The N$_F$-forming molecules are rod-shaped with a substantial longitudinal dipole moment, and self-organize in the N$_F$ phase to yield a spontaneously-adopted, macroscopically uniform, equilibrium ground state polarization field *P*(*r*) of magnitude in the range 2 < *P* < to 10µC/cm$^2$, and which saturates at low temperature at ~90% of that for perfect polar order of the molecular dipoles [9].  Desired geometries of this vector field in devices, which have *P*(*r*) parallel to the RF driving field, are readily achieved by typical LC surface alignment methods or with applied external potentials of only a few volts [10, 11], a small value because the response of $P(r)$ to applied field in the $N_F$ is as a bulk collective reorientational Goldstone mode [9], in contrast with poling of isotropic phases where the field acts on single molecules or groups of a few.

Any $N_F$ material in such an EEO cell geometry will, by virtue of its polar symmetry, exhibit some level of EEO modulation. Given that the nearly saturated polar order appears to be an essential requirement for the $N_F$ phase to be stable [5,14,15], changes of EEO characteristics in the $N_F$ must come from modifying the nonlinear optical polarizability of the constituent molecules. This opens two main research and development paths toward enhancing $N_F$ EEO: (i) Formulating mixtures of $N_F$ hosts doped with large-$\beta$ dipolar EEO chromophores, as reported recently [12, 13], exploiting the strong polar alignment of isolated dipolar (but not $N_F$ forming) molecules in the $N_F$ (the *solvent poling* effect [10, 11]). (ii) creation of molecules which form the $N_F$ phase, and at the same time have large $\beta$. Here we report the achievement of the latter case.

## Materials and Methods

*Molecular Design* – The design theme of these molecules starts with the nanorod shape of RM734 and DIO and their many (~300) $N_F$-forming homologs [14, 15, 16]. In these molecules, polar aromatic groups, for example (-COO$\phi$-) in RM734, are chained in polar fashion to form a linear molecular [core] which has a net dipole $P_c$ generally along the rod axis: [+ $P_c$ -]. To the ends of this [core] are added similarly oriented dipolar groups which extend and terminate the axial dipolar chain, generating what we term the "$N_F$ motif": {[+ $p_h$ -]–[+ $P_c$ -]–[+ $p_t$ -]} of the molecule { }. These end groups are highlighted specifically due to their critical roles: (i) *terminal dipoles* – The dipoles $p_h$ and $p_t$ ideally bring substantial charge to the *very head* and *tail* of the molecules, like the $H_3C$- and -$NO_2$. groups on the ends of RM734, respectively: *head* {+ $H_3C$ -]–[+ $P_c$ -]–[+ $NO_2$ -]} *tail*. If we consider a pair of rods arranged such that their net dipoles $P = p_h + P_c + p_t$ are in the same direction and their ends are in Tail-to-Head van der Waals contact -$NO_2$}···{$H_3C$-, the parallel terminal dipoles $p_h$ and $p_t$ will associate molecular ends by the electrostatic dipole-dipole attraction of proximate Head-to-Tail dipoles [9]. (ii) *multifunctional electrostatic attraction* – The dipolar ends of RM734 are multi-functional, with two sticky sites (-O)$_2$ on the negative end and three sticky sites (-H)$_3$ on the positive end, giving six distinct H-T dipole pairings. This, coupled with the rotational freedom of the $CH_3$ about the C-O bond, the rotational freedom of the C-O bond about the O-$\phi$ bond, and the orientational freedom of the molecule about its long axis distributing the orientation of the $NO_2$ group about the N-$\phi$ bond, creates a flexible attraction between the two molecules, This leads to a transient H-T of the molecules that favors the formation of the $N_F$ phase [9], such that this RM734 scenario, or similar ones with other dipolar groups, *e.g.*, the $\phi$-$F_3$ group in DIO, are nearly universal features of $N_F$-forming molecules [14, 15]. (iii) *chromophoric structures* – Making molecules which are chromophoric while maintaining their $N_F$-forming tendencies means molecular design constrained by the $N_F$ motif: {[+ $p_h$ -]–[+ ($P_c$) -]–[+ $p_t$ - ]}, i.e., by modifying the core, but where the core ideally maintains the molecular rod shape as much as possible, has a longitudinal dipole moment, and has chemical, thermal, and photo stability. Other constraints often encountered in applying organics for EEO are relaxed in the $N_F$, for example molecules need not be modified for solubility, since any molecule having the $N_F$ motif will be soluble in the $N_F$ phase. The core modification directed toward making the rod-shaped molecular core chromophoric is to introduce conjugation to facilitate electron transport along the core axis. The core modification directed toward increasing $\beta$ is to introduce appropriate electron-donating (D) and electron-accepting (A) groups at the h and t ends of the core, respectively, where these groups are connected by conjugated bridging groups (B). The "EEONF motif" results: $\{[+ \boldsymbol{p}_+ \text{-}]–[+ (D – B – A) \text{-}]–[+ \boldsymbol{p}_- \text{-}]\}$, examples of which are given in Refs. [12, 13].

We have assessed the EEO response of PM618 and PM628, two distinct binary mixtures of chromophoric-dye components, designed on the EEONF motif, wherein each individual component exhibits the $N_F$ phase as a neat material. The binary mixing introduces a eutectic depression of the freezing temperature to yield mixture with the $N_F$ phase at room temperature. Its notable that, like RM734 and DIO [11], these components mix in the $N_F$ phase at all concentrations, yielding $N_F$ phases with 100% chromophore concentration, and saturated polar order (polar order parameter, $p \sim 0.9$), as found in RM734 [9].

An important feature of ferroelectric nematic material is the dynamic behavior of the macroscopic reorientation of $\boldsymbol{n}, \boldsymbol{P}$ by external electric field, which can be described by the following equation [9]

$$\gamma \dot{\theta} = |\boldsymbol{P}(\phi) \times \boldsymbol{E}| = PE \sin \phi (t), \qquad (1)$$

where $\gamma$ is the rotational viscosity (comparable in magnitude to the shear viscosity), $\boldsymbol{E}$ is the electric field, and $\phi(t)$ is the angle between $\boldsymbol{n}, \boldsymbol{P}$ and $\boldsymbol{E}$. Under condition of field reversal, reorientation through $\phi(t)$ ~180° requires the characteristic response time $t_0 \sim \gamma/PE$. Thus, electro-optic experiments can used to estimate the viscosity of the PM618 and PM628 mixtures. We found, with 10 V applied across the 100 nm electrode gap, that $t_0 \sim 1000$ s for glassy PM618 or PM628 at room temperature. In comparison, $t_0 \sim 10^{-4}$s for RM734 under similar conditions [10, 11]. Such measurements enable an estimate of the viscosity of our mixtures to be in the glassy range, $\gamma \sim 10^8$ Pa · s at $T$ = 25 °C. The $N_F$ phases of PM618 and PM628 also exhibit the typical strong temperature dependence of $N_F$ viscosity, such that at $T$ = 50 °C the fluid can be oriented in $t_0 \sim 10$ s with 10 V across the 100 nm electrode gap. This temperature dependence produces a stable viscoelastic polar EEO glass at lower temperatures, and, at the same time, enables an easy way to deposit EEO material to confined EEO device geometries by simply filling in the high temperature fluid state and cooling to the glassy $N_F$ phase for operation. These materials were supplied by Polaris Electro-Optics.

*Quantifying ferroelectric nematic EEO* – There are many established methods for measuring the electro-optic coefficients $r_{ij}$ of EEO polymers [17, 18]. Our experimental setup, inspired by the work of Nahata et al [19], is shown in Fig. 1. In the optical schematic, the two polarizing beam splitters (PBS) act as polarizers with their polarizing direction 90 degrees relative to each other, so that the laser is blocked with no sample between the PBS. By convention, we refer to the first PBS as the polarizer and the second PBS as the analyzer. When the laser passes through birefringent material, such as a liquid crystal, a phase difference between the extraordinary and ordinary light components is generally introduced, leading to a change in the polarization state before the laser passes through the analyzer. For a homogenously aligned slab of material with thickness $t$, the introduced phase difference is $\Delta \psi_0^{DC} = \psi_e - \psi_o = (2\pi/\lambda_0)\Delta nt$, where $\lambda_0 = 1550$ nm is the vacuum wavelength of the laser used in the experiment, $\Delta n = n_e - n_o$ is the birefringence of the material and $d = 5$ μm or 15 μm
is the cell thickness. The intensity passing through the analyzer also depends on the relative angle ($\theta$) between the optical axis of the nematic and the polarizer orientation:

$$I = I_0 \sin^2\left(\frac{\Delta\psi_0^{DC}}{2}\right)\sin^2(2\theta) = I_0 \sin^2\left(\frac{\pi\Delta nt}{\lambda_0}\right)\sin^2(2\theta) \quad (2)$$

This equation motivates the scheme for detecting the linear EEO modulation shown in our schematic. When an AC electric field is applied to the material, small changes in the refractive index alter the birefringence ($\delta\Delta n$) and the corresponding phase differences ($\Delta\psi_P^{AC}$).

$$\delta\Delta n = \delta n_e - \delta n_o = \left(\frac{n_e^3}{2}r_{33} - \frac{n_o^3}{2}r_{13}\right)E_3 \quad (3)$$

$$\Delta\psi_p^{AC} = k_0\delta\Delta nt = \frac{2\pi}{\lambda_0}\left(\frac{n_e^3}{2}r_{33} - \frac{n_o^3}{2}r_{13}\right)E_3 d \quad (4)$$

Here, Eq.2 defines the efficiency coefficients $r_{33}$ and $r_{13}$, axis 3 being the ferroelectric polarization (**P**) orientation of the $N_F$ phase. In the cell, the AC modulation field $E_3 = E_0 \sin(\omega t)$ is designed to be optimally parallel to **P**. As a result, the majority of the phase difference arises from refractive index modulations induced along the polar axis, proportional to the $r_{33}$ parameter. We rewrite the phase difference as $\Delta\psi_p^{AC} = \left(\frac{\pi}{\lambda_0}E_3 d\, n_e^3\right)Ar_{33}$, where the parameter $A = 1 - \frac{n_o^3}{n_e^3}\frac{r_{13}}{r_{33}}$ depends on the linear refractive index of the material and the ratio $r_{13}/r_{33}$, which has a value close to 1 (see section S5 I the supplement information).

According to Eq. (1), the induced phase difference $\Delta\psi_p^{AC}$ would produce an AC modulation in intensity $I = I_0 + \Delta I^{AC}$ which could be detected with a fast photodetector. In general, the $r_{33}$ parameter of the material could then be calculated from the ratio between the intensity of AC modulation and DC background.

To further simplify the analysis and increase the sensitivity of the system, the azimuthal angle $\theta$ between the polarizer and the optical axis of the material is adjusted to be 45 degrees. A Soleil-Babinet Compensator (SBC) is added after the sample with its optical axis parallel to one of the principal axis of the sample. The SBC induces another controllable DC phase shift $\Delta\psi_1^{DC}$. By adjusting the SBC, the DC intensity is set to be half of the maximum power ($I_{DC} = I_0/2$), where $\Delta\psi_0^{DC} + \Delta\psi_1^{DC} = \frac{(2m+1)}{2}\pi, m \in \mathbb{Z}$. At this DC set point, assuming the AC modulation field is $E_3 = E_0 \sin(\omega t)$, the small AC intensity modulation can be expressed as:

$$I = I_0 \sin^2\left(\frac{\Delta\psi_0^{DC} + \Delta\psi_1^{DC} + \Delta\psi_p^{AC}}{2}\right) = I_0 \sin^2\left(\frac{\pi}{4} + \frac{\Delta\psi_p^{AC}}{2}\right) \quad (5)$$

which can be expanded linearly into the following expression:

$$I \cong \frac{I_0}{2} + \frac{I_0}{2}\Delta\psi_p^{AC} = I_{DC}\left[1 + \frac{I_{AC}}{I_{DC}}\right] = I_{DC}\left[1 + \frac{\pi}{\lambda_0}E_0 d n_e^3 A r_{33} \sin(\omega t)\right], I_{AC} \equiv \frac{I_0}{2}\Delta\psi_P^{AC} \quad (6)$$

Thus, the value of the $r_{33}$ parameter can be calculated from known parameters and variables and the amplitude ratio of AC intensity/DC intensity.

The above derviation assumes that the appied E field is uniform. However, in the experiment geometry, the in-plane electrodes generate a field that is fairly uniform over most of the gap, but increases rapidly in strength and deviates from in-plane orientation, close to the electrodes, indicated in Fig 1B. In addition, the laser beam intensity profile illuminating the LC is nonuniform (Gaussian), such different parts of the laser beam illuminate regions of the sample exposed to different applied fields. A detailed discussion about these effects could be found in Supplement Section 1. According to previous work [19] and our own analysis, as long as the measurement is close the linear set point of $I_{DC} = I_0/2$, the linear dependence of intensity on the phase difference yields the following expression for $r_{33}$:

$$r_{33} = [\frac{I_{AC}}{I_{DC}}]\frac{\lambda_0}{\pi(F_G V_{app}/a)dn_e^3 A}, \qquad (7)$$

The dimensionless number $F_G$ incorporates all of the geometric effects of integration over the nonuniform electric field and Gaussian beam profiles, where the effective driving field $E_0 = \frac{F_G V_{app}}{a}$ could be defined with $F_G$ defined such that the scaling field is the ratio of measured parameters: the AC driving voltage amplitude across the electrode gap $V_{app}$, divided by the electrode gap, $a$. The $r_{33}$ result reported in this paper is based on Eq.6.

In order to characterize $F_G$, we conducted "x-scan" experiments where the laser scans across the electrode gap along the x axis and the AC signal is monitored (Fig 2B). At the focus of the laser with minimum beam waist, the scanned signal agrees well with the theoretical calculations (Supplementary section 2) and shares a "batman" shape profile (Fig 2E), despite slight mismatch around the electrode edges. In the simulation, the $F_G$ parameter is proportional to the convolution of nonuniform field distribution and Gaussian beam profile. When the beam waist is small compared to the 100 μm electrode gap width, the profile is dominated by the "batman"-shape field distribution which peaks near the electrode edge. On the other hand, when the beam waist is comparable to the gap width, the non-uniformity of the field distribution is smoothed out and the profile resembles the Gaussian beam intensity profile (Fig 2F). The experimental x-scan profiles in Figure 2F with different waists demonstrate this phenomenon. All the experimental measurements (Figure 3,4) are conducted with the beam at the middle of electrode gap with the minimum beam waist, where the polarization field is uniform, and the electrode edge effects are negligible. The corresponding $F_G$ parameters obtained from numerical simulation(SI section 2) are used to determine the $r_{33}$ value, with $F_G = 0.600$ for the 5 μm cell and $F_G = 0.571$ for the 15 μm cell.

*EEO cell* – The EEO cell geometry is the most crucial element in measuring the $r_{33}$ coefficient. As shown in Eq. 6, the cell thicknesses $d$, the electrode gap $a$, the overall alignment of the ferroelectric nematic material, the homogeneity of the electrodes and cell layout (uniform electrode gap and cell thickness) have a direct impact on the measured $r_{33}$ value. To achieve control and consistency over the cell parameters, the experiment is conducted using commercial cells provided by Instec, Inc.. As demonstrated in Fig. 1B, the cell is made up of two glass plates sandwiched together, with two separate rectangular ITO electrodes on the bottom glass surface. The ITO electrodes are separated by a gap $a = 100$μm (red circled region), or $a = 1\ mm$ (black circled region). When a voltage difference is applied across the electrodes, an in-plane field is established in the electrode gap. Exposed regions of ITO on both sides of the cell allow connection to the external circuit. In the experiment, a special cell holder with built-in electrical connector (Instec LCH-S11) is used to hold the sample cell in place and apply AC/DC voltage. The cell thickness is either $d = 5\ \mu m$, or $d = 15\ \mu m$ which is well controlled with spacers and adhesive. The thickness is homogeneous across the whole cell, as confirmed by the uniform birefringence color of liquid crystal in the nematic phase. On both top and bottom glass plates, a thin polymer alignment layer is deposited for orientation control, and the homogenous buffing direction is along the field. To achieve higher modulation field strength, the experiment is conducted in the smaller electrode gap is *a* = 100μm.

*Optical Setup* – Following the schematic in Figure 1A, the polarizing beam splitters (PBS$_1$ & PBS$_2$), sample holder, and Soleil-Babinet Compensator (SBC) form the basic EEO setup. A half-wave plate (HWP) is installed before the first PBS to control the total power of laser light incident on the sample. To minimize the influence of the inhomogeneous applied electric field, which is stronger close to the electrode edges, as shown in Fig. 2, the beam is focused to a small size with $w_0 = 8.5 \ \mu m$ (Figure S3). The small beam size provides an EEO signal which minimizes the effects of field inhomogeneity. In order to achieve tighter focusing, the collimator in the setup expands the beam to 7 mm in diameter, and the lens ($f = 50$ mm) focuses down the beam onto the sample holder with diameter $\sim 20 \ \mu m$. After the sample, the beam is recollimated using the same type of lens and collected into a fast fiber-coupled photodetector with the same type of collimator. The sample holder sits on an three-dimensional (x, y, z) translational stage for fine positional adjustment.

*Experimental procedure* – The ferroelectric nematic material is filled into the commercial liquid crystal cell (EEO cell) at the clearing temperature via capillary force. The cell is then transferred to a polarized light microscope equipped with a hot stage held at 50°C. A 100 V DC voltage is applied to the ferroelectric nematic material at 50°C for 5-10 min for a complete aligning process. The whole process is recorded with the camera attached to the microscope. Subsequently, the cell is cooled down to room temperature with 100 V DC applied. The alignment of the room temperature cell is checked again under polarized light microscope (PLM). Afterwards, the cell is quickly transferred to sample holder where a 30 V DC voltage is applied during the measurement to maintain alignment. Aligned cells were also used in measurement of the refractive indices $n_e$ and $n_o$, as discussed in the Supplement.

In the EEO experiment, at the start, a 1 MHz sinusoidal AC voltage with 10 V amplitude is applied during beam alignment process. The sample is moved to align the beam with the center of the electrode gap based on the *x*-dependence of the Pockels signal(Fig 2E), where the sample cell is translated in *x* direction so that the laser beam scans across the electrode gap (Fig 2B). The compensator (SBC) is adjusted to minimize the total output intensity ($I \sim 0$) for a quick birefringence measurement. Then the SBC is adjusted to the half maximum power set point ($I = I_0/2$), which concludes the aligning process.

Once the setup is ready, several measurements are conducted. (1) In the frequency scan experiment, the frequency of the applied AC voltage is increased from 40 kHz to 10 MHz with fixed 10 V amplitude. For each frequency, the AC signal amplitude $R$ and phase $\theta$ is recorded with the lock-in amplifier, and the DC signal $V_{DC}$ is measured on the oscilloscope. The laser is blocked after each measurement to obtain background readings of the lock-in amplifier $R_{bg}$ and $\theta_{bg}$. The frequency scan experiment is conducted five times in a row to test the repeatability of the setup. (2) In the E field linearity experiment, the frequency scan experiment is conducted from 100 kHz to 3 MHz with the AC amplitude increased from 1 V to 10 V. (3) In the intensity linearity experiment, for each frequency scan from 100 kHz to 3 MHz at 10 V amplitude, the half-wave plate (HWP) is adjusted to control the incident light intensity $I_0$ which is determined by the DC signal $V_{DC}$. (4) The cells are also moved up and down along the gap to test the homogeneity of the cell in the *y* direction. The beam alignment process is carried out each time the cell is moved. The measured r$_{33}$ values were used to verify the reproducibility of the experimental procedure. (5) To confirm the linearity dependence of Pockels signal on the cell thickness, similar experiments are conducted in both $d = 5$ μm and $d = 15$ μm cells for PM618. After the EEO experiment, the sample cell alignment is examined under PLM again to confirm the alignment of N$_F$ is maintained through the experiment. (Fig. S7)

Values of refractive index $n_e$ and $n_o$ as well as the ratio $r_{13}/r_{33}$ are needed for the r$_{33}$ calculation. We measure the refractive index of the PM618 and PM628 material in the aligned EEO cell by studying the multi-beam interference with UV/VIS spectrometer (Supplementary Section S3). The results are in accord with the birefringence measurement in this experimental setup. The $r_{13}$/$r_{33}$ ratio is determined with the dichroism ratio R which is also measured in the aligned sample with spectrometer. The details of these two measurements are provided in the supplement.

## Results and discussion

*Optical characterization of EEO cell alignment and lifetime* – Fig. 2 shows that our EEO cells exhibited quality alignment in PLM, with uniform birefringence color and good extinction with crossed analyzer and polarizer [Note that the N$_F$ materials used here absorbs green/blue light, transmitting red]. The samples also exhibit homogenous dichroism colors in polarized light without an analyzer, with the color of the cell determined by polarization direction and absorbance, the electrode gap appearing darker red when the long axis of material is parallel to the incident polarization of light (Fig 2 C), versus lighter orange when they are perpendicular(Fig 2 D). Further details are discussed in supplementary section 4.

*r$_{33}$ values of PM618 and PM628* – r$_{33}$ values obtained for our glassy ferroelectric nematic mixtures at room temperature are presented in Fig. 4, with, for example, *r*$_{33}$ = 32.8 ± 0.6 pm/V for PM618, and *r*$_{33}$ = 26.0 ± 0.3 pm/V for PM628, at frequency *f* = 100 kHz, and only a weak dependence on frequency in the range 30k-10MHz. To examine the frequency dependence of these *r*$_{33}$ values, we conducted a full frequency scan from 40 kHz to 10 MHz on PM618 and PM628. These r$_{33}$ values are comparable to that of solid-state workhorse material LiNbO$_3$.

As demonstrated in the upper part of Fig. 4, the r$_{33}$ values of PM618 and PM628 only show a weak frequency dependence.in the f$_{low}$ and f$_{mid}$ range, the slope of r$_{33}$ curve in the log-log plot $|\alpha| < 0.08$, which indicates a power law dependence $r_{33} \sim f^{-\alpha}, \alpha < 0.08$. The absence of $f^{-1}$ frequency dependence proves that the intensity-modulated signal in the experiment is not contaminated by the molecular reorientation contribution. The details of frequency dependence analysis are shown in supplementary section S8.

A series of measurements were conducted to provide detailed tests on the setup and further verification of the proposed calculation of r$_{33}$ value (Eq. 7): (*i*) a full frequency scan of the EEO signal and background noise from 40 kHz to 10 MHz (Fig. 4), (*ii*) a linearity check on the electric field and laser intensity (Fig. 3), (*iii*) a repeatability and reproducibility study on the system (Supplementary Fig. S9), and (*iv*) complete x-scan tests on field factor F$_G$ with different beam waists, where the sample is translated stepwise along the z axis through the focus of the laser (Fig. 2 F,G). In order to probe the influence of cell thickness, *r*$_{33}$ of PM618 is measured respectively in a 5μm and 15μm cells. The setup is also tested with a LiNbO$_3$ crystal (Fig S10).In general, the tests of the experimental design showed good agreement with the proposed model. For example, it is expected that the lock-in amplifier output V$_{AC}$ be proportional to the AC modulating field strength $E_0 = V_{app}/a$ and to the DC laser power intensity $I_0$, which is characterized by the DC signal V$_{DC}$. In the experiment, a frequency scan over f$_{mid}$ range is conducted with alternating applied driving voltage *V*$_{app}$ and various DC laser power by rotating the half wave plate. The V$_{AC}$ is plotted versus V$_{app}$ or V$_{DC}$, and the data is fitted linearly at each frequency. As shown in Fig. 3, the signal shows good linearity vs. applied voltage and DC laser power, which is maintained through all the f$_{mid}$ frequency range, with typical results at 1 MHz plotted in Fig. 3. An unpredicted observation is that the linear dependence of *V*$_{AC}$ vs. DC laser power dependence intersects the x axis at a positive offset $I_{\text{offset}}$. When the AC modulation field is turned off, the optical system could achieve good extinction by adjusting SBC with the minimum $V_{DC} < 1\ mV$. When the field is on, it is found out that the SBC cannot perfectly extinguish the DC laser output with the minimum $V_{DC} \sim 5 - 10\ mV$. The exact cause of this phenomenon is under investigation; however, we suspect that there may be additional walk-off between the ordinary and the extraordinary beam. Since the AC modulating signal comes from the interference between the ordinary and the extraordinary beam, the walk-off of two beams causes part of the beam to no longer participate in interference and generates AC signal which act as an offset $I_{\text{offset}}$. As for data analysis, the corresponding $V_{\text{offset}}$ is subtracted out at each frequency $V'_{DC} = V_{DC} - V_{\text{offset}}$ for $r_{33}$ calculation.

The experimental setup demonstrates repeatability and reasonably good reproducibility, and the sample cell also shows good homogeneity along the electrode gap (y direction). Supplementary Fig. S9A, shows five consecutive frequency scan measurements conducted without changes in the setup or movement of the sample cell. The standard deviation $\sigma$ of the $r_{33}$ value is plotted as the error bar, $\sigma$ value is tiny for these measurements, with $\sigma < 0.3$ pm/V indicating a high level of repeatability. The homogeneity is tested by making measurement while moving cell along y axis, The $r_{33}$ value is plotted in Fig. S9B with the standard deviation as the error bar. The $r_{33}$ curves in A and B are nearly identical, while the error bars in Fig.S9B are a bit bigger with $\sigma < 0.8$ pm/V. The increase in the $\sigma$ comes from the slight inhomogeneity in the electrode gap, as confirmed by the microscopy observations. The alignment process could also introduce variation in the signal. With inhomogeneity and reproducibility factors considered, the measurements still provide reasonable accurate $r_{33}$ value.

$r_{33}$ values of PM618 measured in the 5 μm and the 15 μm cells agree well (Fig 3,4), with the $r_{33}$ values of the 5 μm cell higher by ~1.5 pm/V. We suggest that this difference comes from the superior alignment in the thinner cell.

*β-value estimate for mixtures PM618 and PM628*

For a multicomponent chromophoric mixture in a N_F phase, $r_{33}$ is related to the molecular hyperpolarizabilities $\beta^i_{333}$ [13] by:

$$r_{33} = \sum_{i=1}^{n} \frac{c^i f_0 f_\omega^2 \beta^i_{333} \langle \cos^3(\theta) \rangle}{\epsilon_0 n_e^4 v_m^i} \quad (8)$$

Where, $c^i$ is the concentration of the $i$ component, $n_e$ is the extraordinary refractive index of the mixture, $v_m^i$ is the molecular volume of the $i$ component, $f_0$ and $f_\omega$ are separately the local field corrections at DC and optical frequencies. $\langle \cos^3(\theta) \rangle$ is a third-rank orientational order parameter and θ represents the polar angle distribution of the molecules relative to the symmetry axis. For our binary mixtures, the two components have similar structures and concentrations ~50%, so we could estimate a β value under the assumption that $c^i, \beta^i_{333}, v_m^i, f_0, f_\omega$ and $< \cos^3(\theta^i) >$ are same for both components, and then use:

$$\beta_{333} = \frac{r_{33} \epsilon_0 n_e^4 v_m}{f_0 f_\omega^2 \langle \cos^3(\theta) \rangle} \quad (9)$$

Taking $n_e \sim 2$, $f_0 f_\omega^2 \sim 3$, $\langle \cos^3(\theta) \rangle \sim 0.75$, $v_m \sim 69$ nm³ for PM618 and $v_m \sim 68$ nm³ and using the $r_{33}$ value at 100 kHz, we have $\beta_{PM618} = 1.4 \times 10^{-48}$ C³m³J⁻² for PM618 and $\beta_{PM628} = 1.1 \times 10^{-48}$ C³m³J⁻² for PM628.

We did a Gaussian simulation of the hyperpolarizability of the components of the binary mixtures PM618 and PM628 in Turbomole at the CAM-B3LYP/SVPD level of the theory. For PM618, we have $c^1 = 52\%$, $c^2 = 48\%$, $\beta_{333}^1 = 8.26 \times 10^{-49}$ C³m³J⁻², $\beta_{333}^2 = 1.16 \times 10^{-48}$ C³m³J⁻², $v_m^1 = 69.6$ nm³ and $v_m^2 = 69.3$ nm³. For PM628, we have $c^1 = 48\%$, $c^2 = 52\%$, $\beta_{333}^1 = 1.22 \times 10^{-48}$ C³m³J⁻², $\beta_{333}^2 = 1.61 \times 10^{-48}$ C³m³J⁻², $v_m^1 = 72.3$ nm³ and $v_m^2 = 63.7$ nm³. The theoretical $r'_{33}$ value could be calculated based on eqn. 8. We have $r'_{33} = 22.6$ pm/V for PM618 and $r'_{33} = 33.7$ pm/V for PM628. This theoretical $r'_{33}$ value agrees in bulk with measured $r_{33}$ value, the difference could arise from the uncertainty in local field factor $f_0 f_\omega^2$ and orientational order parameter $\langle \cos^3(\theta) \rangle$.

# Conclusion

We have proposed and actualized a design principle that introduces nonlinear optical and EEO effects with enhanced capabilities into molecules that are also capable of forming ferroelectric nematic phases as mixtures of N_F materials. This exciting new dimension in the polar ordering phenomenology of the ferroelectric nematic realm, is exhibited directly in a series of precision measurements of EEO coefficient $r_{33}$, which for the N_F molecules studied is comparable to that of the widely-used solid state EEO material lithium niobate. This favorable comparison of first generation N_F materials with lithium niobate is a significant benchmark, given that N_FS in EEO applications offer several distinct advantages: in processibility, the liquid crystallinity facilitating the fabrication of EO geometries; in stability, intrinsic to its equilibrium polar ordering and the potential for polar glass formation; and in its future potential, based on the plethora of organic molecular chromophoric species having first hyperpolarizability $\beta$ values much larger than those employed here.

# Acknowledgements

This work was supported by the Soft Materials Research Center under NSF MRSEC Grant DMR-1420736, by NSF Condensed Matter Physics Grant DMR-2005170, by the State of Colorado OEDIT Grant APP-354288, and by a grant from Polaris Electro-Optics.

# Conflict of interest

In accordance with University of Colorado policy and our ethical obligations as researchers, we are reporting that several authors at the University of Colorado have a financial interest in Polaris Electro-Optics, Inc., who may be affected by the research reported in this paper. We have disclosed those interests fully to the University and have in place an approved plan for managing any potential conflicts arising from that involvement.

# Author Contributions

XC, CP, MG, and NC conceptualized the project. XC, AB and CP performed the experiments. MG developed and ran computer modeling code. XC, NC and MG analyzed and interpreted the data. XC, AB, MG and JM wrote the paper.

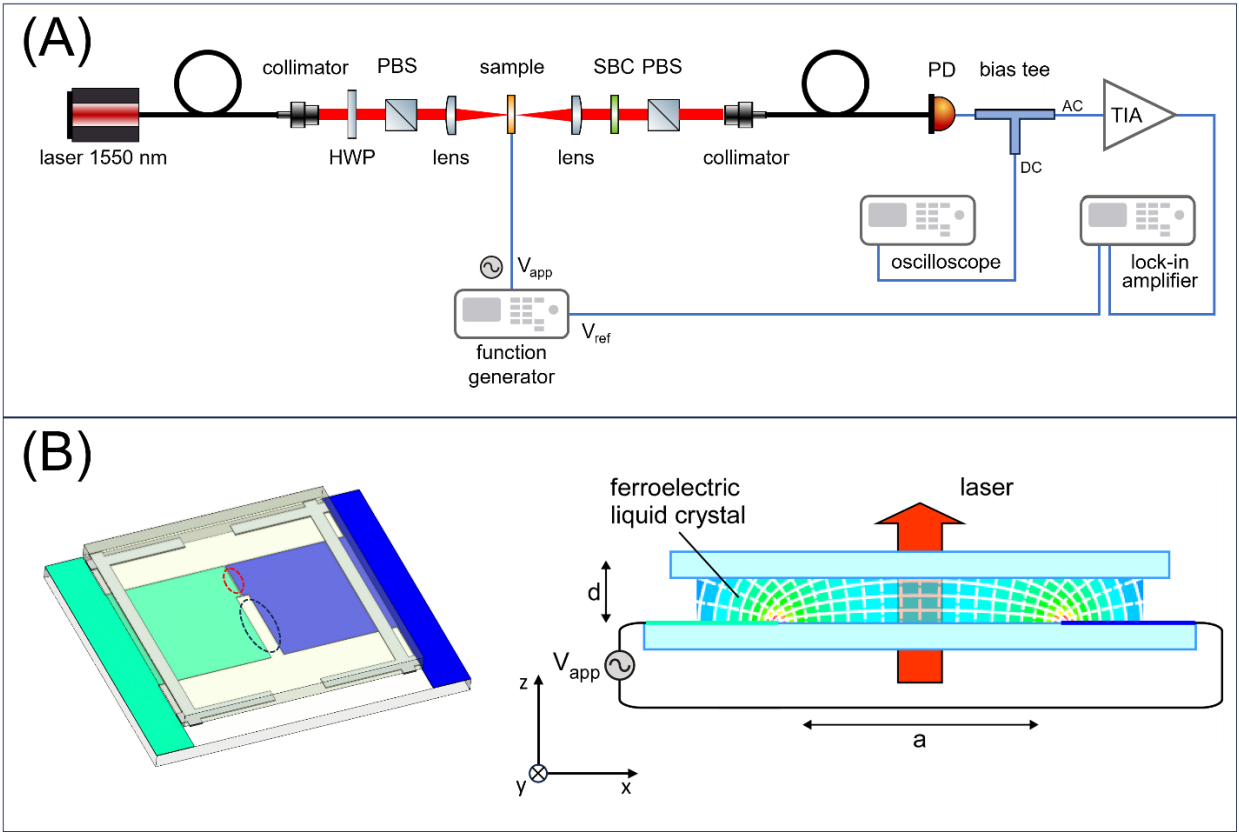

Figure 1: Schematic of the experiment and geometry of the EEO cell.

(A) The optical schematic. The fiber coupled 1550 nm laser is collimated (collimator$_1$) and directed through a half wave plate (HWP) followed by a polarized beam splitter (PBS$_1$). HWP and BSP$_1$ together function as laser power and polarization control. The beam is then focused (lens$_1$) onto the EEO cell which sits on a special sample holder. The sample holder provides an AC sinusoidal driving voltage to the EEO cell with amplitude of $\pm\ 10\ V$ and frequency range from 30 kHz to 10 MHz on top of 30 V DC voltage bias generated by a function generator and home-made DC bias box. The sample holder could be translated on x,y,z axes. The beam after the sample is recollimated with a second lens (lens$_2$), then passed through a Soleil-Babinet Compensator (SBC) for a controlled phase shift between the two polarizations. The beam then passes through a second PBS (PBS$_2$) which has it principal axis rotated 90 degrees relative to the first one (PBS$_1$). Finally, the beam is collected with an identical collimator (collimator$_2$) and fiber coupled to a fast photodetector (PD). The electric signal from the photodetector is first passing through a bias T box to separate AC and DC signal. The DC signal goes to an oscilloscope with 50ohm effective load. The AC signal passes through a transimpedance amplifier (TIA) and then measured with a lock-in amplifier.

(B) The EEO cell geometry. To achieve better control over the sample, the ferroelectric nematic material is filled into a commercial cell from Instec, Inc. The cell consists of two glass slides spaced by $d = 15\ \mu m$. Two ITO electrodes (cyan and blue) are deposited on the bottom glass with a gap between them, and the gap dimension is either $a = 100\ \mu m$ (red circle) or $a = 1\ mm$ (black circle). Alignment layers are deposited on both surfaces with buffing direction along the x axis. When a voltage difference is established between the electrodes, an in-plane field (E) is built up in the gap which would align the ferroelectric nematic polarization (P) into a mostly uniform planar orientation. As depicted by the equal potential and field lines between the electrodes, the field is mostly uniform planar along the x axis in the mid of the gap. However, near the electrode edges, the field grows stronger and deviates from in-plane orientation. The laser beam (red arrow) passes through the cell between the electrode gaps such that the relative phase of its two polarizations is phase modulated by the applied AC field. In order to minimize the influence of inhomogeneity of the field, the laser is focused down to a waist of 8.5 µm passing through the center of the electrode gap. We conduct x-scan experiment (Fig 6) and corresponding theory, simulation (supplement section 1 and 2) for a more accurate measurement of $r_{33}$ value. Noted that the field lines depicted [20] is calculated for thin in-plane electrodes in air as a qualitatively illustration.

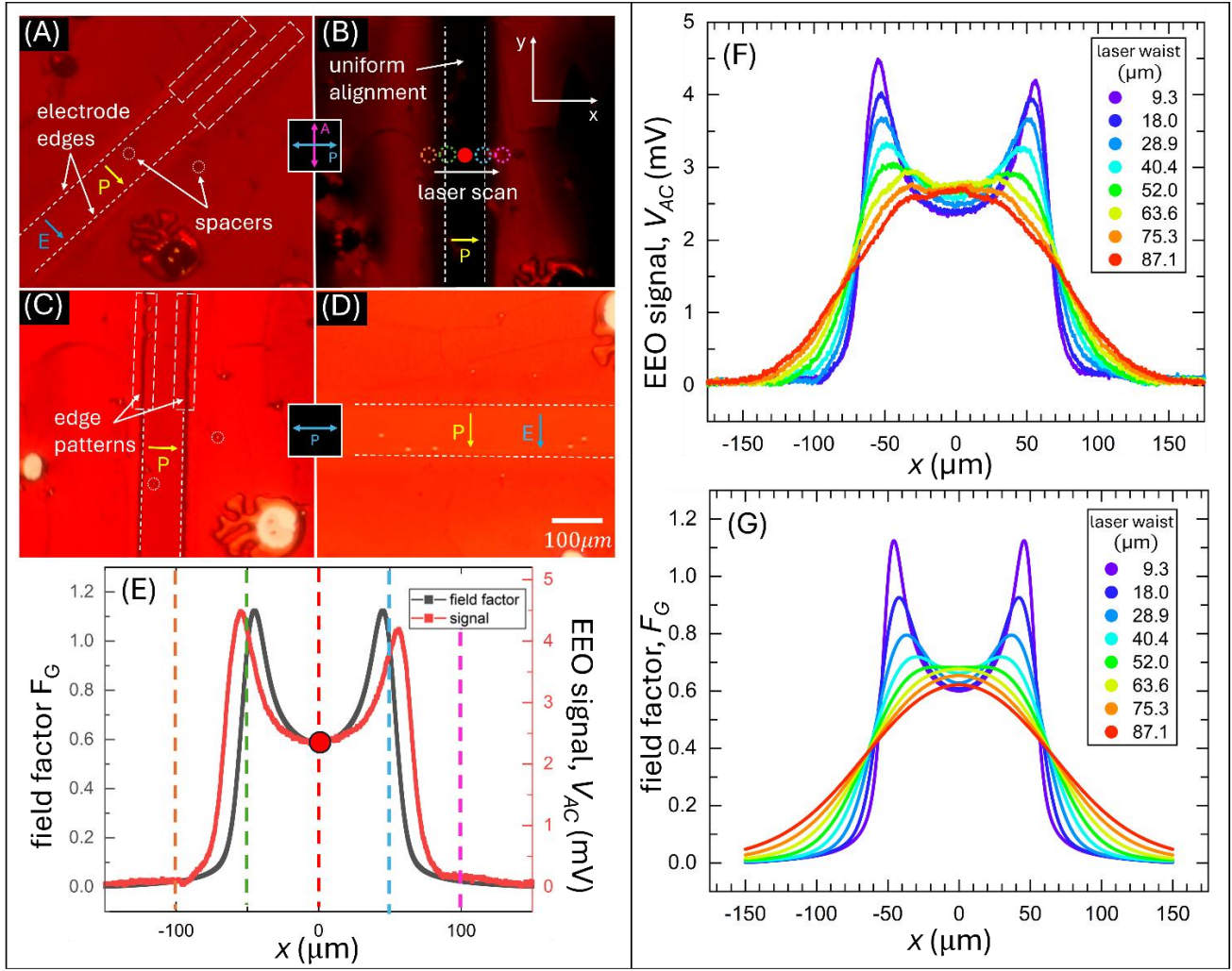

Figure 2: (A-D) Polarized light microscope images of a well aligned EEO cell (A,B) With crossed analyzer and polarizer, the well-aligned EEO cell exhibits homogenous birefringence color and good extinction. (C,D) With only polarizer, the well-aligned EEO cell demonstrates dichroism. The white circles highlight the spacers merged in the material, the uniform alignment is disturbed around the spacers, and those regions are avoided in the EEO experiments. The white dotted line represents the edge of electrodes. The white box signifies the dark lines at the edge of electrodes. Those edge patterns deviate from uniform alignment, and show no symmetry on two electrodes, leading to a mismatch of the x-scan signal between experiment

and simulation near the electrode. The colorful circles in (B) illustrate the different location of laser beam. Their corresponding EEO signal in the x-scan is shown as the dotted line with same color in (E).

(E) The experimental x-scan of EEO signal and the calculated field factor from simulation. The details of simulations are explained in the Supplementary Information. The experimental data and theoretical simulation agree generally well without any fitting involved. The EEO signal peaks near the electrode edges due to the stronger field strength there. However, this "batman" shape signal is not as symmetric as predicted in the simulation because of non-symmetrical edge pattern of liquid crystal. The mismatch in the x scale could be a result of the edge pattern structure. Despite these differences, the similarity of the shapes of the data and simulation curves enable the field factor $F_G$ at the middle of the gap (orange point) where the polarization is uniformly aligned and the laser illuminates, to be determined for use in the calculation of $r_{33}$.

(F,G) X-scan curves of EEO signal with PM618 in a $d$ = 5 μm cell with different gaussian beam waists, (F) experiment (G) theoretical calculation. The EEO signal is recorded when the sample cell translated parallel to the 100 μm electrode gap along the x axis. The EEO signal profile is determined by the convolution of inhomogeneous field of in-plane electrodes and the gaussian beam profile. When the beam waist is much smaller than the electrode gap width, the profile resembles the 'batman' in-plane field distribution where the field strength peaks near the electrode edge. On the other hand, when the beam waist is comparable to the electrode gap width, the profile resembles the gaussian intensity distribution of the focused laser. Different beam waists are achieved by scanning the sample cell through the focus of the laser in z axis. For simplicity, we demonstrate half of the scan here, the full scan could be found in the supplement (Figure S6 S8) The theoretical EEO signal profile is characterized by the field factor $F_G(x)$. Using the exact same laser waist, the theoretical profile agrees generally well with experimental results without any fitting involved which further validates our model and calculation for r33 value. The detail of the theoretical calculation is explained in the supplement.

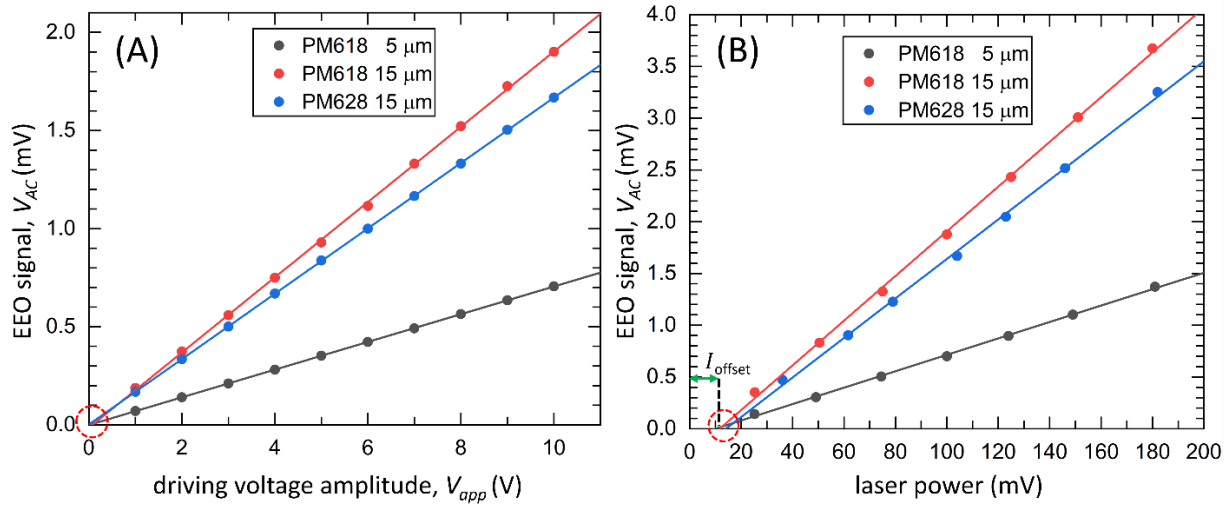

Figure 3: AC modulation signal $V_{AC}$ at $f = 1$ MHz. $(A)$ $V_{AC}$. vs. driving AC voltage amplitude $V_{app}$. (B) and DC laser intensity . From equation 5, the AC modulation signal is expected to be proportional to both field strength and DC laser power. To confirm the effect, the r33 frequency scan (100 kHz-3 MHz, f_mid) of PM618 and PM628 cells are conducted with various AC modulating voltage (1-10 V) and various DC laser power which is characterized by the DC signal $V_{DC}$ (20 mV-180 mV). At each frequency, the processed EEO

signals from lock-in amplifier depend linearly on applied voltage and laser power. The fitting at 1 MHz is chosen to show the typical results. Despite the good linearity in both field and laser power dependence, an important observation is that the fitted lines in field dependence passes through the origin (red circle), while the fitted line in laser power has a constant offset on the x axis $I_{\text{offset}}$. We suspect the offset relates to the beam walk-off caused by an inhomogeneous phase mask generated by the EEO (see the main text for more detail). When calculating $r_{33}$, at each frequency, the corresponding $I_{\text{offset}}$ is subtracted from the $V_{\text{DC}}$.

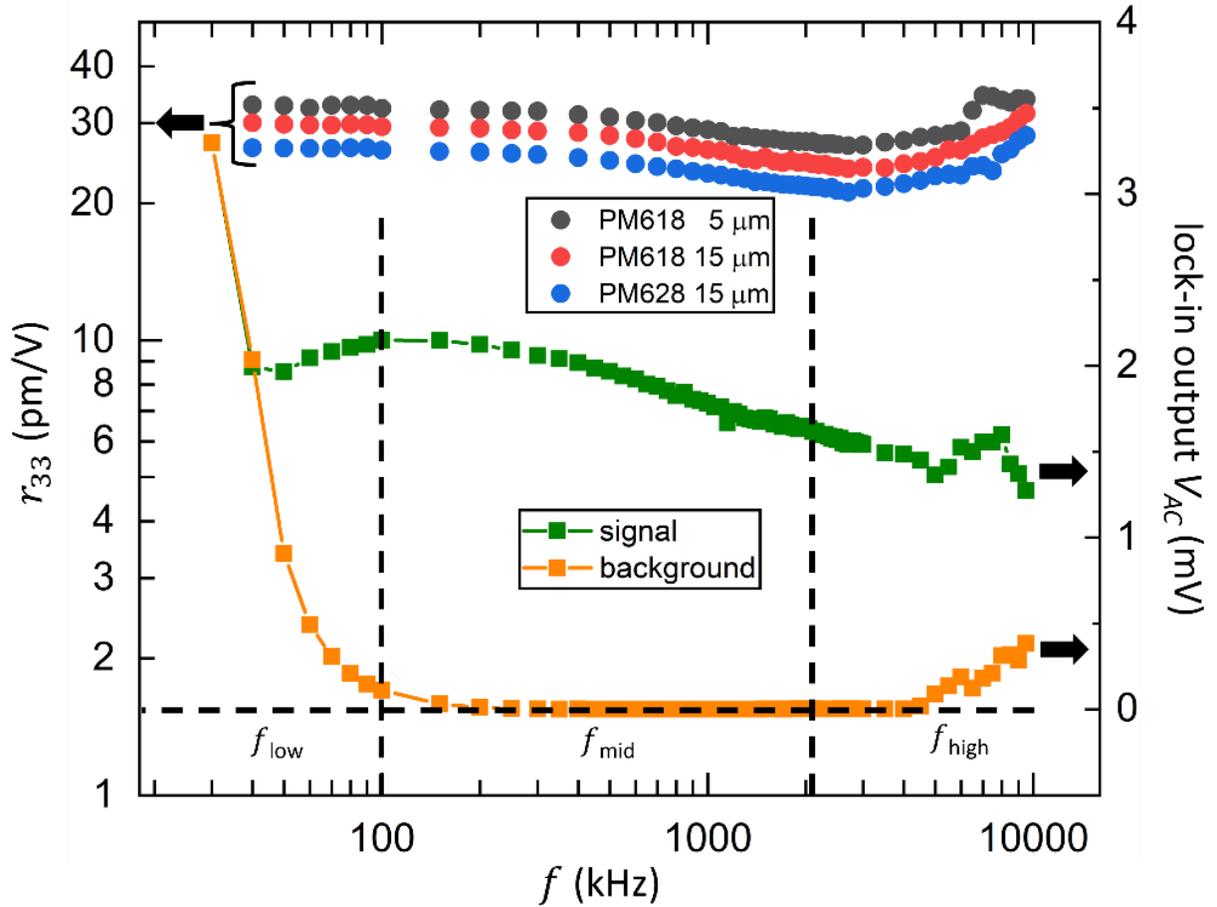

Figure 4: The full frequency scan of $r_{33}$ value of PM618 and PM628 at room temperature (top) and the typical signal vs background over the experimental frequency range. The PM618 mixture is examined in a 5 μm and 15 μm cell where the PM628 is examined only in a 15 μm cell. The frequency range of the detection is from 40 kHz to 10 MHz. As shown in the top of the figure, the measured $r_{33}$ value for PM618 and PM628 is around 30 pm/V with a weak frequency dependence, discussed in the main text. The lack of dependence on frequency indicates that the measured $r_{33}$ originates from the electronic EEO rather than from molecular reorientation which would generate $r_{33} \propto 1/f$. At the bottom of the figure, the typical frequency dependence of signal and background are plotted. The signal refers to the lock-in amplifier amplitude output $V_{AC}$ (R channel) which measures the electronic EEO modulated light signal. The background refers to the same lock-in amplifier output $V_{AC}$ when the laser is blocked, which measures all the electronic background sourced from radiation or equipment crosstalk. The frequency dependence of the signal is the main cause of the corresponding $r_{33}$ frequency dependence. The background could be sorted into low (40 kHz-100 kHz, $f_{low}$), middle (100 kHz-3 MHz, $f_{mid}$) and high (3 MHz-10 MHz, $f_{high}$) frequency ranges. In the $f_{mid}$ range, the background is negligible with typical S/N>400. In the $f_{low}$ range, even though the background is comparable to signal, the background has a stable amplitude and phase which could be easily subtracted out (see main text). In the $f_{high}$ range, the background rises up quickly and cannot be easily subtracted out. We are focusing on the $f_{mid}$ range in this experiment.

**Supplementary Information:**

**Fully-chromophoric Ferroelectric Nematics for Electronic Electro-optics**

*Xi Chen[1], Andrew Bradfield[1], Chirag Patel[2], Pavel Savechenkov[2], Jason W. Sickler[2], Gianlorenzo Masini[2], Joseph E. Maclennan[1], Noel A. Clark[1], Cory Pecinovsky[2]\*, Matthew A. Glaser[1]\**

[1]Department of Physics, University of Colorado Boulder, Boulder, CO 80309
[2]Polaris Electro-Optics, Inc., Broomfield, CO 80020

\*corresponding authors: matthew.glaser@colorado.edu, cory@polariseo.com

## Section S1: The Field Factor F_G

In this section, we discuss the effects of non-uniform electric field and laser intensity on the EEO measurements. We use the sample coordinate system shown in Fig. 1, where z is along the thickness of cell, y is parallel to the electrodes, and x is across the electrode gap, parallel to the average direction of the in-plane applied electric field.

Polarized light microscope observations of sample cells both before and after the Pockels effect experiments were carried out confirm that a uniform, homogeneous alignment is maintained in most of the region between the electrodes where the $r_{33}$ measurements are made (see Fig. 2 and Fig. S8 below). Near the electrode edges, the alignment is slightly non-uniform, with the cell displaying complex textures whose appearance depends on the liquid crystal material. The cell alignment is discussed further in Section S4 below. For the purposes of calculating the electric field distribution, however, we assume that the entire liquid crystal sample is homogeneously aligned in the DC field.

In a time-varying applied electric field $\boldsymbol{E} = \boldsymbol{E}(x,y,z)\sin(\omega t)$, the extraordinary and ordinary refractive indices $n_e$ and $n_o$ of the ferroelectric liquid crystal vary because of the Pockels effect as

$$n_e(x,y,z,t) = n_e^0 + \frac{n_e^3}{2} r_{33} E_x(x,y,z) \sin(\omega t) \tag{S1}$$

$$n_o(x,y,z,t) = n_o^0 + \frac{n_o^3}{2} r_{13} E_x(x,y,z) \sin(\omega t), \tag{S2}$$

where $E_x(x,y,z)$ is the E-field component parallel to the liquid crystal polarization **P**, and $n_e^0$, $n_o^0$ are the refractive indices in the absence of an applied field. The corresponding phase shift induced in an optical beam propagating through a thin slab of the liquid crystal of thickness d$z$ is:

$$d(\Delta\psi(x,y,z)) = k_0 \Delta n \, dz = \frac{2\pi}{\lambda_0} \Delta n^0 dz + \frac{\pi}{\lambda_0} (n_e^3 r_{33} - n_o^3 r_{13}) E_x(x,y,z) \sin(\omega t) dz. \tag{S3}$$

Here $\Delta n^0$ is the sample birefringence in the absence of applied field. Assuming that the liquid crystal is uniformly aligned through the thickness of the cell $d$, the total phase difference introduced by the liquid crystal is then simply

$$\Delta\psi(x,y) = \frac{2\pi}{\lambda_0} \Delta n^0 d + \frac{\pi}{\lambda_0} (n_e^3 r_{33} - n_o^3 r_{13}) \sin(\omega t) \int_0^d E_x(x,y,z) \, dz. \tag{S4}$$

In our experiments, the Pockels cells were either $d = 5$ μm or $d = 15$ μm thick and the waist of the focused beam was $w_0 = 8.5$ μm. The corresponding Raleigh length is $z_0 = \pi w_0^2 / \lambda_0 = 146$ μm in vacuum, or about 292 μm in a liquid crystal with $n_e = 2$. The cells may therefore be regarded as optically thin and any beam walk-off can be neglected.

Under these conditions, interference between the ordinary and extraordinary components of the transmitted beam occurs independently at locations in the sample exposed to the beam. Consequently, for each point after the analyzer, the transmitted intensity varies about the DC set point $I_{DC} = I_0/2$ as:

$$I(x,y) \cong \frac{I_0(x,y)}{2} + I_0(x,y)\Delta\psi_p^{AC}(x,y) = \frac{I_0(x,y)}{2} + \frac{\pi I_0(x,y)}{2\lambda_0} n_e^3 A r_{33} \sin(\omega t) \int_0^d E_x(x,y,z) dz, \tag{S5}$$

where $I_0(x,y)$ is the profile of the Gaussian beam incident on the liquid crystal cell. The total optical power measured at the detector will have both DC and AC components:

$$P_{\text{DC}} = \iint dxdy\, I_0(x,y)/2 \tag{S6}$$

$$P_{\text{AC}} = \frac{\pi}{2\lambda_0} n_e^3 A r_{33} \sin(\omega t) \iiint dxdydz\, I_0(x,y) E_x(x,y,z) \tag{S7}$$

In our experimental setup, the electrodes are much longer than their separation. The liquid crystal alignment is generally good everywhere along the electrodes (Figs. 2, S8), with any inhomogeneities producing only small variations in the EEO signal ($\lesssim 3\%$). For this calculation, we can therefore reasonably assume translational symmetry along the y axis, so that the electric field does not depend on y. The Gaussian beam-weighted field-correction factor $F_G$ may then be defined as

$$F_G(x_0) = \frac{1}{d} \frac{\iiint dx\, dy\, dz\, I_0(x-x_0,y) F_x(x,z)}{\iint dx\, dy\, I_0(x-x_0,y)}, \tag{S8}$$

where $F_x(x,z) = E_x(x,z)/E_0$ is the scaled applied electric field, $d$ is the thickness of the Pockels cell, and $E_0 = V_{\text{app}}/a$ is the nominal field strength given by the ratio of the applied driving voltage $V_{\text{app}}$ and the electrode gap $a$.

This field-correction factor, which accounts for the effects of the Gaussian beam shape and the inhomogeneity of the electric field generated by the in-plane electrodes, is a key element in accurately determining the magnitude of $r_{33}$. A numerical calculation of the field factor was carried out for a laser beam with waist $\omega_0$ centered at position $x_0$ in the electrode gap. With all other parameters fixed, the form of $F_G(x_0)$ resembles the AC signal profile when the laser is scanned across the electrode gap (along the x direction). We used such x-scan experiments to probe the agreement between the experiment and theory. Scans were performed with different laser waists by translating the 5 μm PM618 cell parallel to the light path (along the z axis), through the focus of the laser. The results are shown in Figs. 2F, S1 and S2. The waist of the Gaussian beam was measured using the knife-edge method (Fig. S3). As explained in the **Methods** section of the paper, the EEO signal profile along x is given by the convolution of the (non-uniform) electric field distribution with the Gaussian beam profile. When the beam waist is small compared to the 100 μm electrode gap, the observed EEO profile has a "batman" shape, with peaks near the electrode edges. When the beam waist is comparable to the size of the gap, on the other hand, the spatial non-uniformities in the applied field are smoothed out and the measured intensity profile resembles that of the Gaussian beam. This effect is illustrated for different beam waists in the x-scan profiles of Fig. 2F. The theoretical $F_G$ profiles calculated with these beam waists are plotted for comparison in Fig. 2G. The measured experiment profile shapes are described quite well by the simulation. However, the observed intensity profiles are systematically narrower than predicted, resembling simulation curves calculated with a 30% smaller waist value. For example, the 40.4 μm experimental curve in Fig. 2F matches the simulation with a 28.9 μm waist in Fig. 2G. This discrepancy could be caused by the presence of the "edge patterns" observed in the microscope. In the simulations, the polarization in the electrode gap is assumed to be perfectly aligned and uniform. In practice, the director/polarization field often forms complex, non-uniform textures near the electrodes (the "edge patterns"). This non-uniformity of the liquid crystal alignment could result in a moderate reduction of the applied field strength near the electrode edges, an effect qualitatively equivalent to a reduction in the beam waist. This mismatch has little impact on our calculation of $r_{33}$, however, since all the experimental measurements (Figs. 3,4) are conducted in the middle of electrode gap, where the polarization field is uniform and any electrode edge effects are negligible, using the smallest beam waist (i.e., at the laser focus). Despite the discrepancy between the simulation and experimental results near the electrode edges shown in Fig. 2E, they match well in the middle of the gap, inspiring confidence in the faithfulness of the computed field factor values $F_G$ used to calculate the magnitude of $r_{33}$. At the middle of the electrode gap, we found $F_G = 0.600$ for the 5 μm cell and $F_G = 0.571$ for the 15 μm cell.

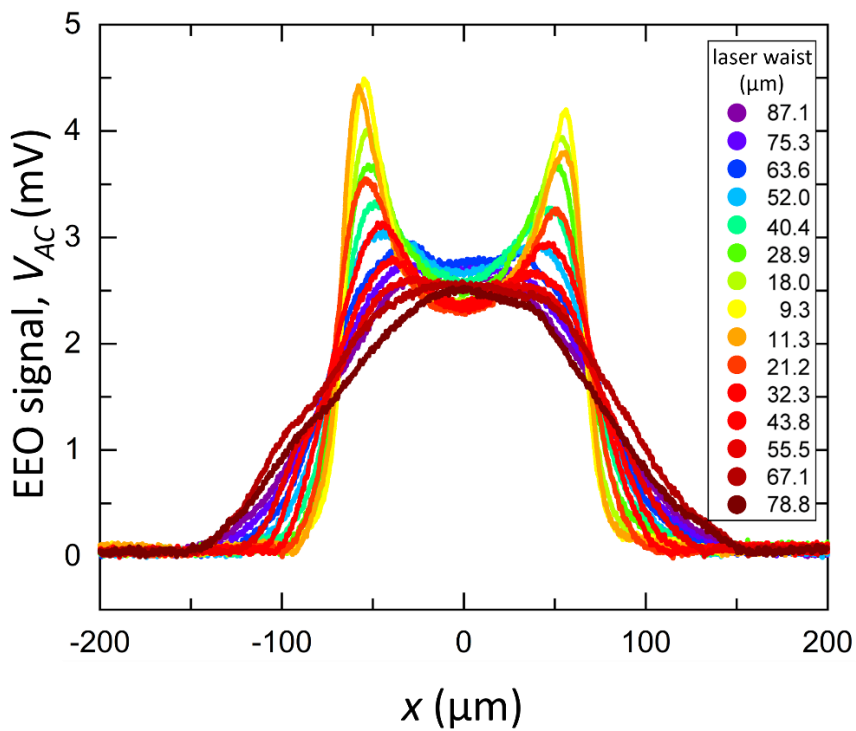

**Figure S1**. Optical signal vs. lateral position in a 5 $\mu m$ thick PM618 cell at different locations along the laser beam. The electrode gap is 100 μm (the electrode edge locations are indicated by vertical dashed lines). Each scan is essentially a convolution of the laser beam profile with the applied field strength at each location in the cell.

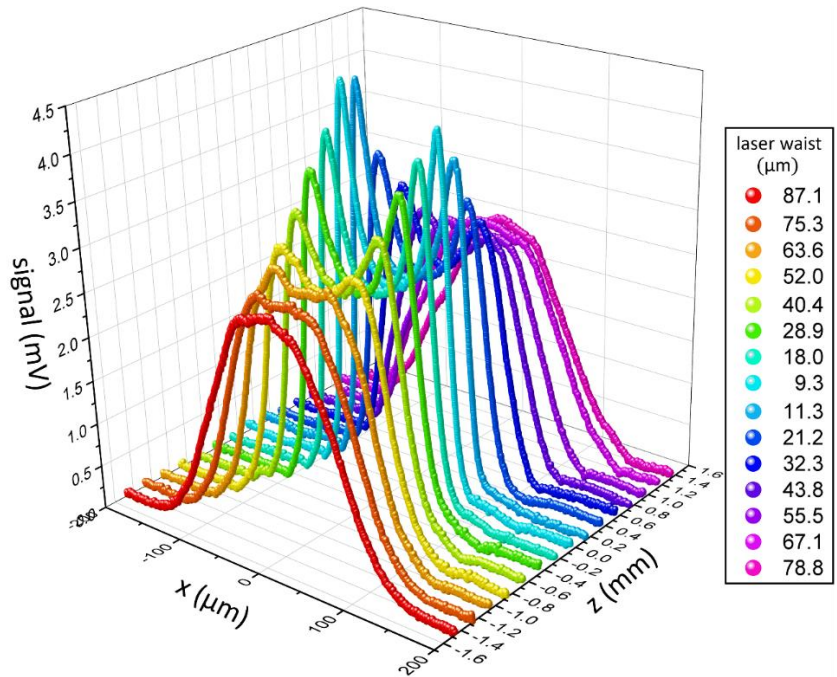

**Figure S2**. Optical signal vs. lateral position x between the electrodes in a 5 μm thick PM618 cell as the sample is moved along the laser beam, through the focus at z=0. The data are the same as in Figure S1.

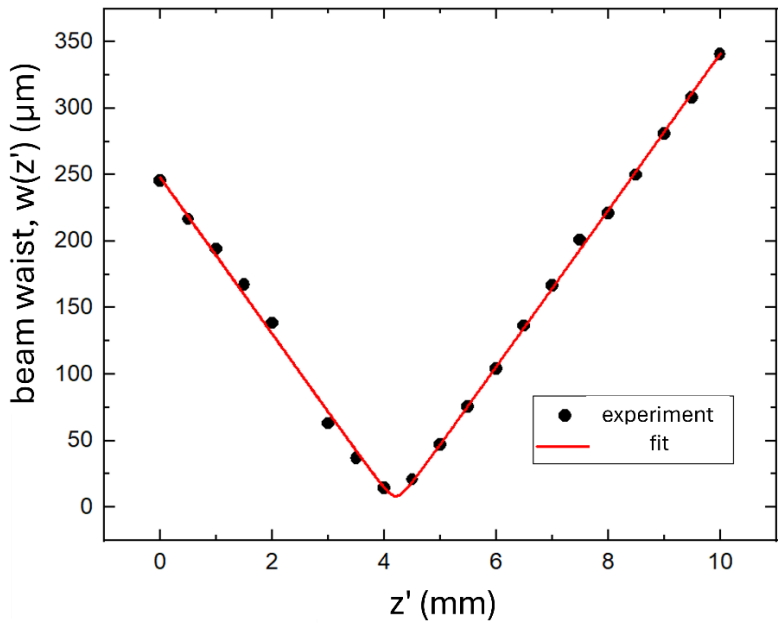

**Figure S3**. Beam waist vs. position along the beam. Knife-edge measurements were conducted to determine the beam width as a function of distance along the focused beam. These measurements were fitted to the expression for the waist of a Gaussian beam $w(z') = w_0\sqrt{1 + (z'/z_R)^2}$, where $z_R = \pi w_0^2/\lambda$, with the result shown in the graph. The fitted curve was used to calculate the laser waist at arbitrary cell locations z in subsequent experiments, with $z = 0$ corresponding to the position with minimum beam waist (the laser focus). Since the estimated Rayleigh length of the beam (~150 μm in air) is much larger than the thickness of the cells used in the experiment, it is reasonable to assume that the divergence of the beam is negligible and that the beam waist is constant through the thickness of the sample.

## Section S2: Field Factor Calculation

In the previous section, we explained how the effects of the non-uniform electric field and Gaussian beam distribution are accounted for by the field factor $F_G$. To improve the accuracy of the measured $r_{33}$ values, a detailed simulation was carried out to obtain the exact electric field distribution in the liquid crystal cell. The field factor $F_G$ was then calculated by convolving the Gaussian beam profile with the calculated electric field.

The cell geometry, with the ferroelectric nematic liquid crystal (5 or 15 μm thick) sandwiched between two sheets of soda-lime glass, is sketched in Fig. S4A. In the experiments, the glass is 1 mm thick, much thicker than the liquid crystal layer, so that it may be taken in the calculation to extend to infinity in both positive and negative z directions. Two 50 nm-thick ITO electrodes separated by a 100 μm gap are located on the bottom glass plate. The polyimide alignment layers are not included in the simulation.

In order to carry out the calculation, one needs to know the dielectric constants of all of the component materials, including the ferroelectric nematic. In the simulation, we assume that the liquid crystal polarization is perfectly uniformly aligned between the electrodes. While polarized light microscope observations reveal that in many cells there is some non-uniformity in the texture near the electrode edges, the polarization is in fact homogeneously aligned across most of the gap, so the assumption of a uniform polarization field seems a reasonable starting point.

We solved the generalized Poisson equation,

$$\nabla \cdot \left[ \epsilon_r(\boldsymbol{r}) \nabla \phi(\boldsymbol{r}) = -\frac{1}{\epsilon_0} \rho(\boldsymbol{r}) \right], \tag{S9}$$

where $\epsilon_r(\boldsymbol{r})$ is the relative permittivity (in general a second-rank tensor), $\phi(\boldsymbol{r})$ is the electrostatic potential, and $\rho(\boldsymbol{r})$ is the free-charge density. In principle, $\rho(\boldsymbol{r})$ should also include the bound-charge density arising from divergence of the spontaneous polarization density, but we assume a highly uniform ferroelectric nematic layer with negligible bound-charge density and containing no free charges, so $\rho(\boldsymbol{r}) = 0$ everywhere. A finite-difference approximation method is used to solve the generalized Poisson equation in 2D by successive over-relaxation [1] within a 300 μm x 300 μm domain with a grid spacing of Δ = 0.05 μm. We impose Neumann boundary conditions (zero normal component of electric field) on the outer boundaries of the computational domain and Dirichlet boundary conditions on the electrodes to maintain a fixed potential difference V between the electrodes. Dielectric anisotropy is ignored and $\epsilon_r(\boldsymbol{r})$ is taken to be a scalar quantity. The applied electric field was simulated in both 5 μm and 15 μm cells, with results for the latter shown in Fig. S4B.

We now consider an incident beam with a Gaussian intensity profile:

$$I_0(x, y) = \frac{1}{2\pi w_0^2} \exp\left[-\frac{x^2 + y^2}{2w_0^2}\right] \tag{S10}$$

The Gaussian-averaged effective field factor $F_G$ may be computed by convolving the time-dependent electric field with the intensity profile:

$$F_G(x_0) = \frac{1}{d} \frac{\iiint dx\,dy\,dz\, I_0(x - x_0, y) F_x(x, z)}{\iint dx\,dy\, I_0(x - x_0, y)}, \tag{S11}$$

where $x_0$ is the center position of the Gaussian beam, $d$ is the thickness of the Pockels cell, and $F_x(x,z) = E_x(x,z)/E_0$, where $E_0 = V_{app}/a$ is a nominal field strength corresponding to the ratio of the applied driving voltage $V_{app}$ and the electrode gap $a$. The integrals over x and y are carried out from $-\infty$ to $+\infty$, and the integral in z over the thickness of the cell.

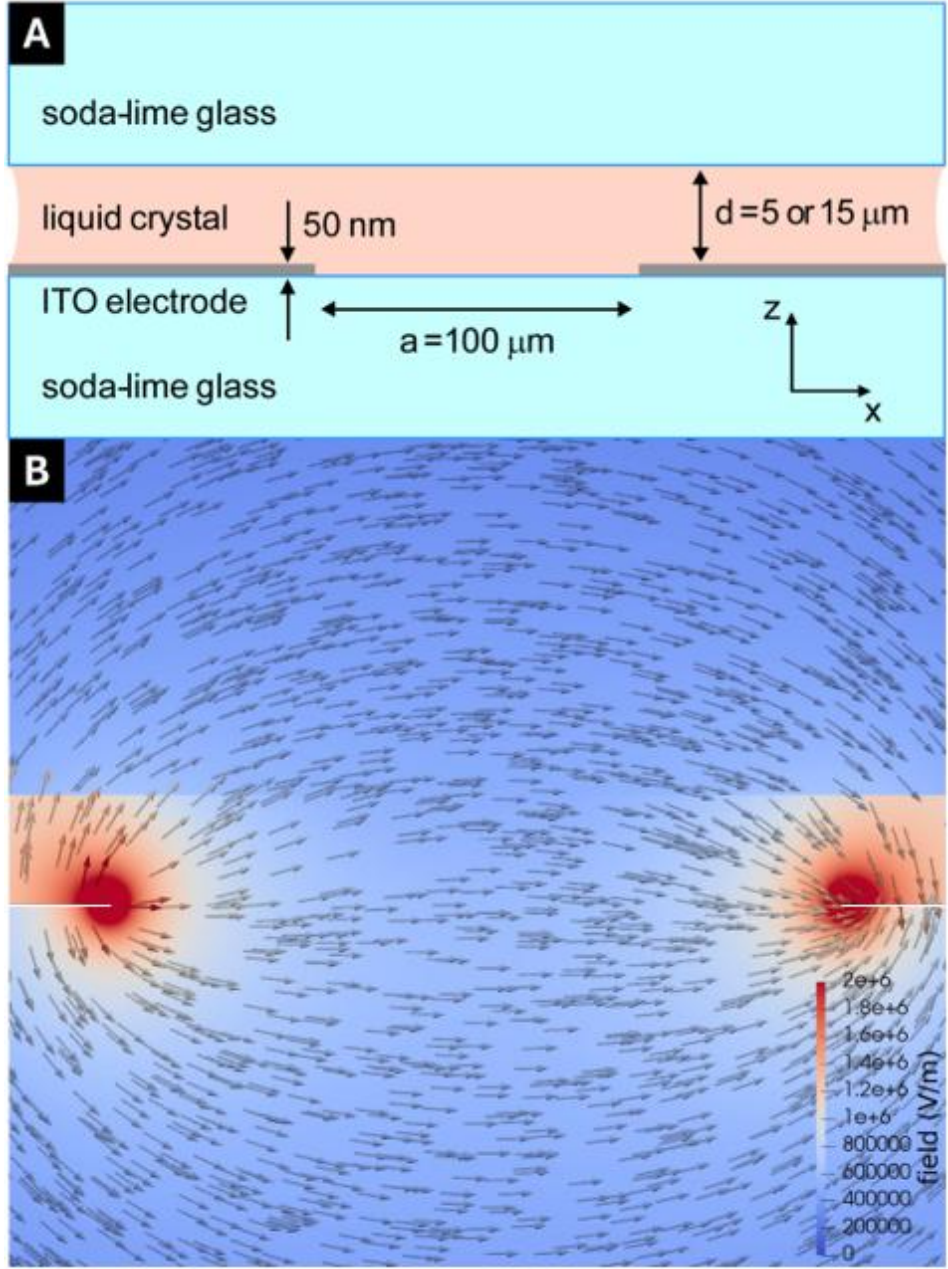

**Figure S4**. (A) Pockels cell geometry, showing soda-lime glass substrates (cyan), a thin film of ferroelectric nematic liquid crystal (pink), and ITO electrodes (dark gray). (B) Computed applied electric field strength in a 15 μm cell. The field strength rises quickly near the electrode edges but is relatively uniform over a large region between the electrodes where the experiments are performed.

As shown in Fig. 2E, the calculated field factor profiles generally reproduce the "batman ear" shapes seen experimentally in transverse scans of the liquid crystal cell. Since the electric field strength and polarization orientation are essentially constant near the middle of the gap, the field factor at the center of the gap is only weakly dependent on the beam width when the beam waist is small ($w_0 \lesssim 15\ \mu m$).

In general, despite the slight discrepancy between the model and experimental profiles pointed out above, the simulation should provide accurate $F_G$ values in the middle of the gap, where most of the measurements used for calculating the $r_{33}$ values reported here were made.

**Section S3: Refractive Index Measurement**

*Introduction* – The polarized optical transmission spectra of uniformly aligned, rubbed cells with a thickness of 15 μm were measured at vacuum wavelengths $\lambda$ corresponding to vacuum wavevectors $k = 2\pi/\lambda$ in the range 3.1 μm$^{-1}$ < $k$ < 6.2 μm$^{-1}$, using an Agilent Cary 5000 spectrophotometer. The extraordinary and ordinary refractive indices were probed in turn by aligning the incident polarization respectively parallel and normal to the liquid crystal director. The light intensity transmitted through the cell, $I(k)$, exhibits thin film interference [2, 3], which is constructive at wavevectors $k_N$ locating maxima in $I(k)$:

$$k_N = N * \left(\frac{2\pi}{2d * n(k)}\right), \tag{S12}$$

where $N \geqslant 0$ is the integer order of the peak, $d$ is the film thickness, and $n(k)$ is the refractive index of the sample [3, 4, 5]. For the lowest order peak (N = 1), the phase shift due to the path difference between the directly transmitted and doubly reflected light is $2\pi = 2d\ k_1\ n(k_1)$. Since at longer wavelengths the refractive index $n \sim 2$, we expect $k_1 \sim 2\pi/(4d) \approx 0.1$ μm$^{-1}$ and $\lambda_1 \sim 4d = 60$ μm for the first order.

The transmission spectrum of the mixture PM628 at T = 25°C with the optical polarization set parallel to the director to measure $n_e$ is shown in Fig. S5. The raw spectral data points and a fit in the form of a sinusoid with varying amplitude are plotted. If in the wavelength range of interest $n(k_N) = n$ is constant, then we have $k_N \propto N$ and $\Delta k_N$, the spacing in $k$ between successive maxima [4, 5], is given by:

$$\Delta k_N = (k_N - k_{N-1}) = \frac{\pi}{nd}, \tag{S13}$$

which is also constant. The fractional spacing between the interference maxima is then

$$\frac{\Delta k_N}{k_N} = \frac{k_N - k_{N-1}}{k_N} = \frac{1}{N}. \tag{S14}$$

The transmission spectrum of the mixture PM628 at T = 25°C with the optical polarization set parallel to the director to measure $n_e$ is shown in Fig. S5. The raw spectral data points and a fit in the form of a sinusoid with varying amplitude fringe spacing are plotted, showing a disposition of the interference maxima with $n(k_N)$ changing by only 0.5% per order, indicating an index only weakly dependent on $k$. In this case the corresponding index $n(k_N)$ can be calculated from Eq. S13.

Taking $d$ = 15 μm, $\lambda_N$ = 1.55 μm, $k_N = 2\pi/\lambda_N$ = 4.053 μm$^{-1}$, and $n \sim 2$ (as estimated from the typical neighboring peak spacing ($\Delta N$ = 1) in Fig. S5), the estimated range of the orders of the interference fringes observed in our measurements is ~26 < N < ~56. Obtaining exact N values for the peaks in Fig. S5 would require knowing $n(k)$ over the spectral range (0.1 μm$^{-1}$ < $k$ < 3 μm$^{-1}$, 60 μm > $\lambda$ > 2 μm), which is not available. In the absence of precise knowledge of N, we simply number the peaks in Fig. S6 from left to right by index 1 < M < 31.

The refractive index at the wavenumber of each fringe, $n(k_M)$, was determined solely from the peak spacing values $\Delta k_M$ [6], as follows. The positions of the interference maxima were obtained by fitting the spectra over the whole M range to a single function $I(k) = A(k)*\sin^2(B(k)) + C(k)$, where $k = \omega/c$ is the vacuum wavevector, and A, B, and C are polynomials as follows: $A(k) = A_1+A_2*k+A_3*k^2$, $B(k) = B_0+B_1*k+B_2*k^2+B_3*k^3$, and $C(k) = C_0+C_1*k+C_2*k^2+C_3*k^3$. The background term $C(k)$ has been subtracted from the plots in Fig. S5.

The peak positions $k_M$ are determined by the third-order polynomial $B(k)$, which gives an excellent global representation of the phase function, $k_M$ vs. M, of the sinusoid. The fitted peak positions of $I(k)$ computed from $B(k)$ are plotted as $k_M$ vs. M, giving the solid line in Fig. S6A. Peak positions were also independently obtained by performing a local fit around each maximum, giving the red points in Fig. S6A. Since there was no systematic residual deviation between these two sets of $k_M$ vs. M values, as shown in Fig. S6B, the experimental $k_M$ vs. M was taken to be that determined from $B(k)$ (the black line in Fig. S6A). The values of $B_0$ and $B_1$ are both positive, as expected. Taking these terms alone, corresponding to the case of no dispersion (i.e., with $n$ independent of $k$), gives the dashed line in Fig. S6A. The slope of the $B(k)$ curve decreases with increasing $k$, indicating that $n(k)$ does, in fact, increase with $k$, as expected in a dispersive organic medium.

<u>Calculating the refractive index</u> – The intervals $\Delta k_M = (k_M - k_{M-1})$ express the difference in $k$ required to fit one additional wavelength in the path difference 2d in the vicinity of $k \approx k_M$. Since $\Delta k_M$ changes slowly with M, by only ~ 0.3% per M interval, we can assume $n$ to be the local average value in the interval $\Delta k_M$ and use Eq. S13 to calculate this local $n(k)$, which is also changing by only ~0.3% per M. The refractive index dependence on wavenumber is calculated using the peak positions $k_M$ obtained from $B(k)$, with the results shown in Fig. S7.

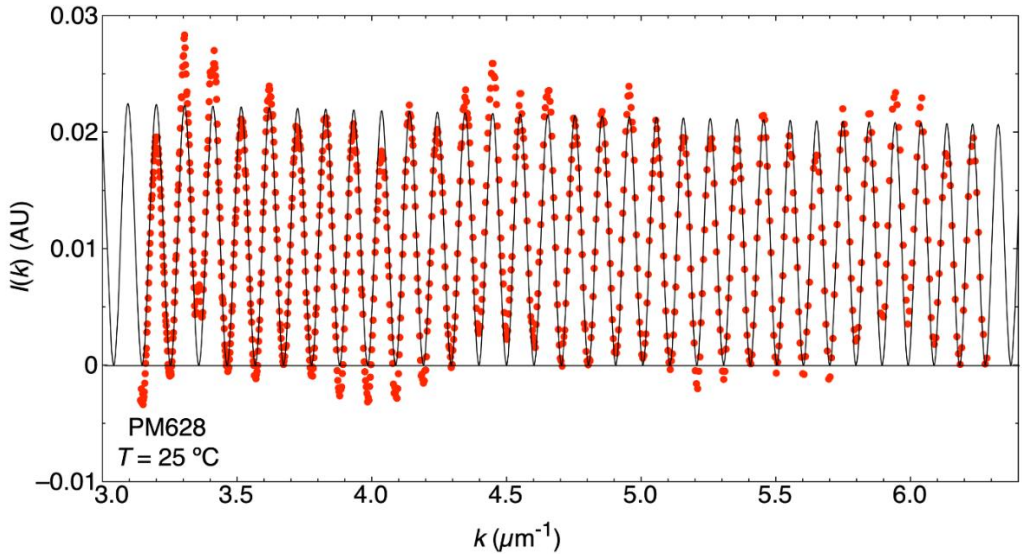

**Figure S5**. Transmitted intensity vs. wavenumber for a PM628 cell with thickness $d$ = 15 μm. Red dots: spectral intensity measured with incident optical polarization parallel to the director, probing $n_e$. Solid Line: intensity function of the form $I(k) = A(k)*\sin^2(B(k)) + C(k)$, where $k = \omega/c$ is the vacuum wavenumber. A, B, and C are polynomials given in the text. The peak locations $k_M$ yielded by $B(k)$ are shown in Figure S6.

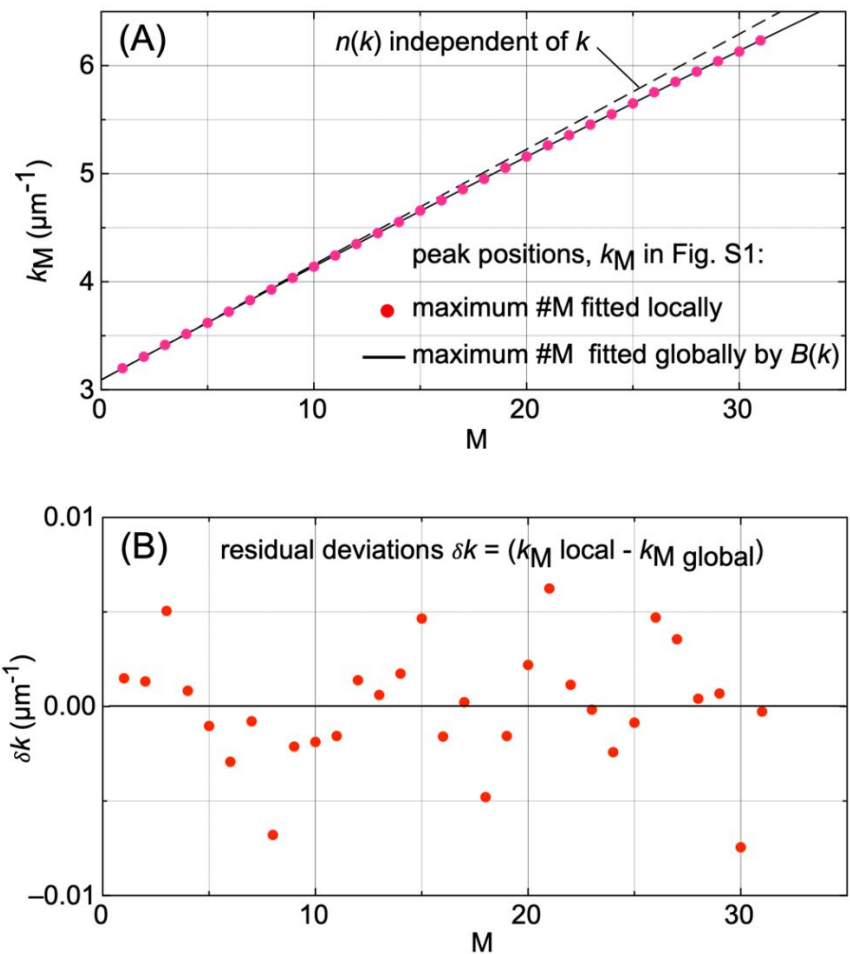

**Figure S6**. (A) Peak wavenumber $k_M$ vs. index $M$ obtained by fitting the transmission spectrum of the PM628 cell shown in Fig. S5. Red dots show the locally fitted locations of the interference maxima of $I(k)$ vs. wavenumber $k$. The solid black line is the global fit to these peak locations obtained from $B(k)$. The dashed line shows the expected behavior in the absence of dispersion of the refractive index. (B) The residual deviations between the global and local fits are random, showing that the global fit describes the phase of $I(k)$ well.

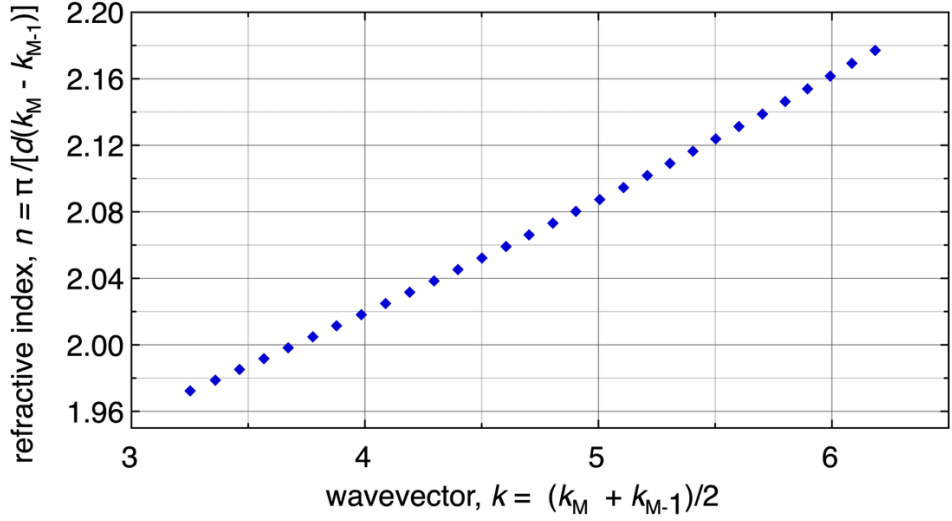

**Figure S7**. Calculated refractive index vs. wavenumber for the PM628 cell. The measurement gives the local average $n(k)$ in each interval $\Delta k_M$, calculated using Eq. S13, which is assigned to $k = (k_M + k_{M-1})/2$, the $k$ value at the center of the interval.

### Section S4: Alignment and Long-term Stability of Glassy $N_F$ Samples

Achieving stable alignment with a high degree of polar order is of key importance in the development of Pockels effect materials for applications. Ferroelectric nematics have some notable advantages in this regard. First, ferroelectric nematics have polar order parameters that are large ($p\sim0.9$) even without poling [7], demonstrating polar order much larger than that typically achieved in poled polymers [8, 9, 10]. Second, the polarization orientation is a Goldstone mode, meaning that every orientation is a ground state. Once the glassy ferroelectric nematic phase is aligned along one direction, its well-ordered lifetime is not limited by relaxation of the polar order but only by the thermodynamic stability of the phase. The glassy ferroelectric nematic EEO cells tested in our setup maintain their alignment and EEO properties for weeks without crystallizing or requiring realignment.

The uniformity of the planar-aligned ferroelectric nematic material in the cell determines how accurately the EEO coefficients can be measured. In our experiments, the cell alignment is monitored and evaluated under a polarized light microscope. Well-aligned EEO cells have a uniform birefringence color (Fig. 2A) and show good extinction (Fig. 2B) between crossed analyzer and polarizer. Such cells also exhibit homogenous dichroism colors in polarized light, where in the absence of an analyzer the color of the cell is determined purely by the absorbance. The ferroelectric nematics studied here absorb green/blue light, appearing dark red when the molecular director (***n***,***P***) is parallel to the polarization of the incident light (Fig. 2C) and light orange when the director is perpendicular (Fig. 2D).

The liquid crystal materials are glassy and highly viscous at room temperature. Heating the samples to 50°C, however, lowers their viscosity enough that a relatively small field of 1 V/$\mu$m is sufficient to reorient the polarization ***P*** homogeneously. If the sample is then cooled to room temperature again, this polarization orientation is locked in. The resultant alignment is maintained for weeks, even in the presence of an AC modulation field. Although the liquid crystal is generally well-aligned within the electrode gap, the molecular director typically deviates from uniform planar alignment near the electrode edges and around spacer balls in the gap. The dark regions along the electrodes seen in some cells, shown in Figs. 2A,C and Fig. S8, exhibit complicated textures we refer to as "edge patterns". These textures are typically different at the two electrodes and show no translational symmetry along y. These non-uniformities in liquid crystal alignment are not accounted for in the simulation of electric field in the cell and are posited as the major reason for the difference between the experimental data and the simulation of the variation of EEO signal strength as the beam is scanned across the electrode gap seen in Fig. 2E. When the probe beam is in the middle of the electrode gap, however, the effects of the non-uniform alignment near the electrodes are negligible: the experimental measurements agree well with the simulations, and the calculated field factor $F_G$ may be used with confidence to determine $r_{33}$ values. Unless otherwise stated, the $r_{33}$ values reported here are all derived from such measurements made at the middle of the electrode gap.

The accuracy of the $r_{33}$ measurement depends on the uniformity of the aligned ferroelectric nematic being maintained over the course of the experiment. We have confirmed that the alignment of the liquid crystal is well preserved throughout measurements performed over several days, with an example of the cell textures before and after EEO measurements were carried out shown in Fig. S8. In fact, the alignment of the director/polarization field in cells of both PM618 and PM628 could be maintained for weeks, the polar order being preserved even in the absence of external poling fields. Loss of alignment does take place if the liquid crystal material crystallizes, which sometimes occurs, for reasons that are not well understood, after many months.

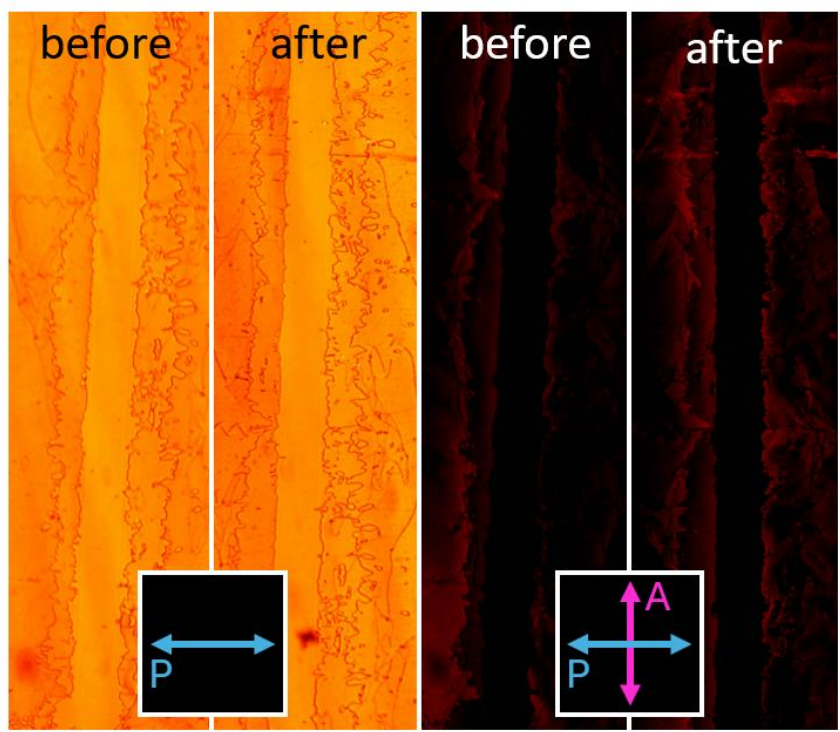

**Figure S8**. Textures observed before and after carrying out EEO measurements on a PM628 cell. The uniformity of the aligned region between the electrodes remains essentially unchanged over the course of the experiment.

## Section S5: The $r_{13}/r_{33}$ Ratio

The value of the $r_{13}/r_{33}$ ratio may be determined using the following equation:

$$\frac{r_{13}}{r_{33}} = \left(\frac{n_e}{n_o}\right)^4 R_x, \quad R_x \equiv \frac{\chi^{(2)}_{113}}{\chi^{(2)}_{333}} = \frac{\langle P_1 \rangle - \langle P_3 \rangle}{3\langle P_1 \rangle + 2\langle P_3 \rangle}, \tag{S15}$$

where $\langle P_1 \rangle$, $\langle P_2 \rangle$ and $\langle P_3 \rangle$ are the averaged Legendre polynomials of the molecular orientational distribution function $f(\cos\theta)$ expressed in terms of the polar angle $\theta$ measured from the polar symmetry axis

$$\langle P_n \rangle = \int_{-1}^{1} P_n(x) f(x) dx, \text{ where } x = \cos\theta. \tag{S16}$$

The precise molecular orientational distribution is not known so for convenience we assume a Gaussian distribution, $f(x) \propto e^{-(1-x)^2/2\sigma^2}$, with $x$ in the range $-1 < x < 1$, characterized by the single parameter $\sigma$. The integration of such a Gaussian distribution between finite limits in $x$ is, in general, algebraically complex, with the result expressed in terms of special functions (incomplete gamma functions). In the $N_F$ materials studied here, however, the order parameter is generally large, indicating that $\sigma$ is small, and the integrals may be evaluated to a good approximation by extending the lower limit of $x$ from $-1$ to $-\infty$. In this case, the normalized distribution is $f(x) = (\alpha/\sigma)e^{-(1-x)^2/2\sigma^2}$, with $\alpha = \sqrt{2/\pi}$, and the Legendre polynomials are given by $\langle P_1 \rangle = 1 - \alpha\sigma$, $\langle P_2 \rangle = 1 - 3\alpha\sigma + (3/2)\sigma^2$, and $\langle P_3 \rangle = 1 - 6\alpha\sigma + (15/2)\sigma^2 - 5\alpha\sigma^3$.

The width $\sigma$ of the Gaussian distribution can be expressed in terms of $\langle P_2 \rangle$ as

$$\sigma = \alpha \left[ 1 - \sqrt{1 - \left(\frac{\pi}{3}\right)(1 - \langle P_2 \rangle)} \right]. \tag{S17}$$

The value of $\langle P_2 \rangle$ can, in turn, be expressed in terms of the dichroic ratio $R$ of the sample as

$$\langle P_2 \rangle = \frac{R - 1}{R + 2}. \tag{S18}$$

The dichroic ratios $R$ of PM618 and PM628 were determined experimentally by measuring the absorbance of light polarized parallel and perpendicular to the director using a spectrometer. Following the formalism outlined above, the $r_{13}/r_{33}$ ratios for PM618 and PM528 were found to be 0.316 and 0.3124 respectively.

## Section S6: Repeatability, Reproducibility, and Homogeneity

Five consecutive frequency scan measurements conducted without changing the experimental setup or moving the sample cell are shown in Fig. S9A. The standard deviations σ of the $r_{33}$ values obtained in these measurements, plotted as error bars, is tiny, with σ < 0.3 pm/V indicating the level of repeatability. In order to test the homogeneity of the ferroelectric alignment, the sample cell is translated parallel to the electrode edges. Each time the sample is repositioned, beam alignment is ensured before a frequency scan is carried out. The $r_{33}$ values are plotted in Figure S9B, again with the standard deviations, a measure of the reproducibility of the measurements, represented by error bars. The $r_{33}$ curves in Figs. S9A and B are nearly identical, although the error bars in Fig. S9B are slightly bigger, with σ < 0.8 pm/V. Microscopy observations suggest that the increase in the spread of the $r_{33}$ values measured at different locations arises from small inhomogeneities in the director alignment in the electrode gap. Minor differences in the beam alignment at each sample position could also result in variations in the signal strength. In summary, the experiments are found to be repeatable and reasonably reproducible, giving us confidence in the measured values of $r_{33}$.

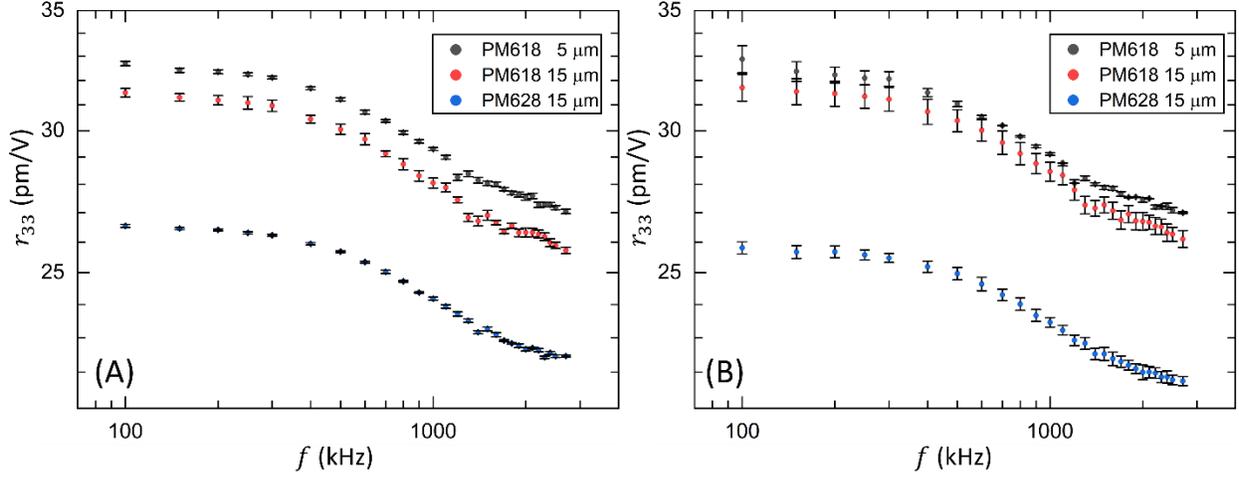

**Figure S9**: Repeatability, reproducibility and homogeneity trials. (A) In the repeatability experiment, five consecutive frequency scans over the $f_{mid}$ range are conducted without any changes to the experimental setup. The standard deviation of the measured $r_{33}$ values, $\sigma$, is shown as error bars. The overall repeatability of the measurements is excellent, with $\sigma < 0.3$ pm/V.  (B) In the reproducibility trial, the sample is translated parallel to the electrode gap (along the y direction), with a full beam alignment carried out each time the sample is moved. The standard deviation of $r_{33}$ is seen to be somewhat bigger than in (A), with $\sigma < 0.8$ pm/V. We propose that inhomogeneities in the alignment at different places along the electrode gap contribute to the observed spread in the measured $r_{33}$ values. The smaller variance in measured $r_{33}$ values in the thinner of the two PM618 cells suggests that the alignment of PM618 is better in the thinner cell than in the thicker one, an inference supported directly by a comparison of the two cells in the polarized light microscope. Similarly, the difference in the variances of the $r_{33}$ values measured in the 15 µm cells of PM628 and PM618 is consistent with the alignment being more uniform in the PM628 cell.

### Section S7: Extracting $r_{33}$ from the Measured Optical Output

In general, the output optical signal is expected to comprise a small, high-frequency AC (30 kHz to 10 MHz) intensity modulation superimposed on a much larger DC signal. A fast InGaAs photodetector with no built-in transimpedance amplifier is used to detect the transmitted intensity. A commercial bias tee separates the AC and DC signals: the DC signal, $V_{DC}$, is measured directly with an oscilloscope with a 50 Ω load, while the AC signal is further amplified using a commercial transimpedance amplifier with an input impedance of 10 kΩ. The amplified AC signal, $V_{AC}$, is then measured using a lock-in amplifier.

The output voltage from the photodetector is given by $V_o = P * R(\lambda_0) * R_{\text{load}}$, where $P = \int I \, dA$ is the total laser power on the detector, $R(\lambda_0)$ the responsivity of the photodetector, and $R_{\text{load}}$ the effective load, which is different for the AC and DC signals. In an ideal situation, assuming a perfectly homogenously aligned sample, uniform light intensity and electric field amplitude, we would have:

$$\frac{V_{\text{AC}}}{V_{\text{DC}}} = \frac{P_{AC} R_{\text{load}}^{\text{AC}}}{P_{DC} R_{\text{load}}^{\text{DC}}} = \frac{I_{AC} R_{\text{load}}^{\text{AC}}}{I_{DC} R_{\text{load}}^{\text{DC}}} = \frac{R_{\text{load}}^{\text{AC}}}{R_{\text{load}}^{\text{DC}}} \frac{\pi}{\lambda_0} E_0 d n_e^3 A r_{33} \qquad (S19)$$

or

$$r_{33} = \frac{V_{AC}}{V_{DC}} \frac{R_{load}^{DC}}{R_{load}^{AC}} \frac{\lambda_0}{\pi E_0 d n_e^3 A} \ . \tag{S20}$$

In practice, both the applied electric field and the laser intensity are spatially non-uniform. This is accounted for by scaling the optical field amplitude by the field correction parameter $F_G$, yielding:

$$r_{33} = \frac{V_{AC}}{V_{DC}} \frac{R_{load}^{DC}}{R_{load}^{AC}} \frac{\lambda_0}{\pi F_G E_0 d n_e^3 A} \ , \tag{S21}$$

where $E_0 = V_{\text{app}}/a$ is the nominal applied field (the ratio between the amplitude $V_{\text{app}}$ of the driving AC voltage and the electrode gap width $a$).

## Section S8: Frequency Dependence of the r₃₃ Signal

In this section, we investigate the possible causes for the apparent frequency dependence of the measured r₃₃ values. As shown in Fig. 4, the lock-in amplifier detects an AC background signal even when the laser is blocked. This could come from imperfect electrical shielding of the signal lines and equipment crosstalk. The background signal shows a strong frequency dependence, rising dramatically at lower frequencies (40 kHz–100 kHz, f_low), largely due to operating near the low-frequency limit of the lock-in amplifier (30 kHz). The background also increases at high frequencies (3 MHz–10 MHz, f_high), due to enhanced electromagnetic radiation between the experimental components. Wrapping the BNC cables in aluminum foil helped to reduce the background. Nevertheless, the remaining background severely degrades the EEO signal at frequencies above about 5 MHz. In the mid-frequency range (100 kHz–3 MHz, f_mid), however, where most of the experiments were performed, the background is negligible, with the S/N ratio typically bigger than 400.

In the f_low region, the background could easily be subtracted out. We define the background output from the lock-in amplifier when the laser is blocked as R_bg and θ_bg, so that $V_{AC}^{bg} = R_{bg} e^{i\theta_{bg}}$, the signal output from the lock-in amplifier when the laser is on as R_s and θ_s, so that $V_{AC}^S = R_S e^{i\theta_s}$. The measured AC signal is the sum of the background and the EEO signal, $V_{AC}^S = V_{AC}^{bg} + V_{AC}^R$. The EEO signal may then be expressed as $V_{AC}^R = V_{AC}^S - V_{AC}^{bg}$, with an amplitude of $|V_{AC}^R| = \sqrt{R_s^2 + R_{bg}^2 - 2 R_s R_{bg} \cos(\theta_s - \theta_{bg})}$. This subtraction method works well in the f_low range, the processed signal $|V_{AC}^R|$ providing accurate r₃₃ values even if the background is as large as the signal. This background subtraction method was validated by manually adjusting the Soleil-Babinet compensator to the next half-maximum power set point ($I = I_0/2$). In this way, an additional $\pi$ phase shift is introduced into the EEO signal $V_{AC}^R$, resulting in a change in the measured signal, $V'^S_{AC}$, from roughly constructive interference to destructive interference between $V_{AC}^R$ and $V_{AC}^{bg}$ or vice versa. Applying the subtraction method to the new measured signal $V'^S_{AC}$ always gave the same real EEO signal, $V_{AC}^R$.

The background in the f_mid range is negligible, so the observed frequency dependence in r₃₃ comes directly from the measured AC signal $V_{AC}^S$. There are three possible causes: (1) The devices (function generators) used to apply the AC and DC voltages to the EEO cell. (2) The detection setup, including the bias tee, the transimpedance amplifier, and the lock-in amplifier. (3) An intrinsic frequency dependence of the electrical properties of the glassy ferroelectric nematic mixture. We monitored the voltage applied to the EEO cell in the experiment with an oscilloscope and observed no frequency dependence, ruling out cause (1). To test cause (2), we kept the detection setup unchanged, except for substituting a function generator for the photodetector. To calibrate the frequency dependence of the detection circuit, a sinusoidal test signal 10 mV in amplitude was applied to the input of the detection circuit and the lock-in amplifier output recorded. A drop in signal strength was observed in the $f_{high}$ range (above 3 MHz) due to approaching the cutoff frequency of the bias tee and the transimpedance amplifier. The measured frequency dependence of the detection system was scaled out of all the data shown in this paper but is anyway negligible in the $f_{mid}$ range. Ultimately, we decided that the observed frequency roll-off of $r_{33}$ in the $f_{mid}$ range is probably a reflection of the intrinsic properties of the glassy ferroelectric nematic material (cause (3)) but resolving this question was beyond the scope of this experiment.

### Section S9: Calibrating the $r_{33}$ Measurement: LiNbO$_3$ Reference Sample

The EEO response of a commercial LiNbO$_3$ crystal (EOL3050X-AR1550 from Newlight Photonics) is used as a reference to calibrate the experimental setup. The crystal has dimensions 3x3x5 mm (x,y,z). The crystal is x-cut, with gold electrodes coating the z-faces. The electric field is applied parallel to the ferroelectric polarization, as in the N$_F$ cells. The x-faces are anti-reflection coated to maximize transmission at 1550 nm. The LiNbO$_3$ crystal is mounted in the EEO system in place of the liquid crystal cell with no further changes to the experimental apparatus. The results are shown in Fig. S10.

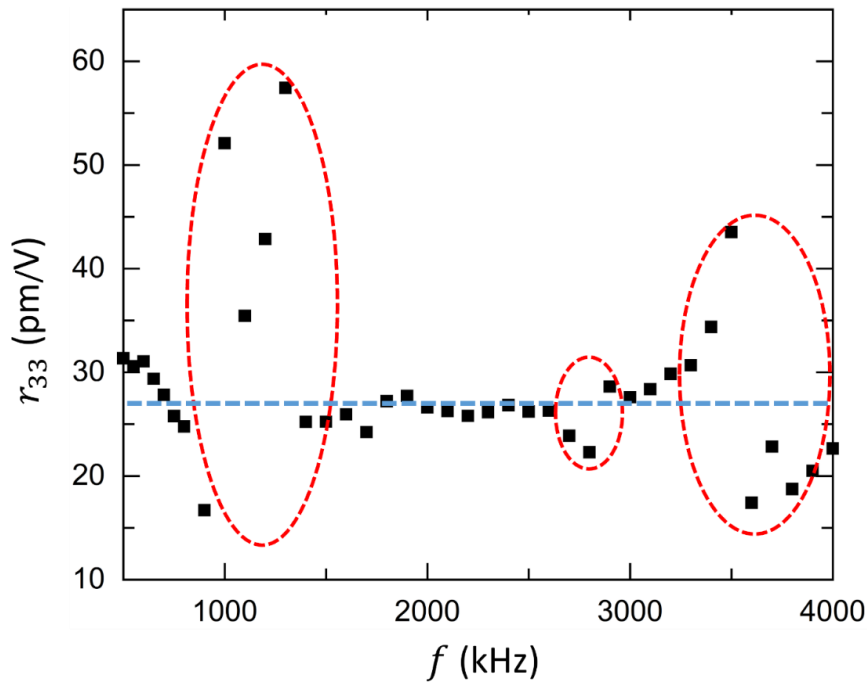

**Figure S10**. Experimental system calibration. The electrooptic coefficient $r_{33}$ of a LiNbO$_3$ reference crystal was measured at frequencies from 500 kHz to 4 MHz. The blue dashed line depicts the averaged measured $r_{33}$ value of around 27 pm/V. This is slightly lower than values for LiNbO$_3$ given in the literature (these are typically around 30 pm/V), a difference we attribute to the uncertainty in the magnitude of the field factor $F_G$ in our experiment. In computing the $r_{33}$ values, we assumed for simplicity a uniform field $V/a$ in the electrode gap, likely an overestimate of the actual field strength that would lead to an underestimate of the $r_{33}$ value. The red ovals highlight the frequency ranges where anomalous signals were recorded. These signals are thought to correspond to piezoelectric resonances of the LiNbO$_3$ crystal.